\documentclass[aps,twocolumn,superscriptaddress]{revtex4-1}

\usepackage{amsmath}
\usepackage{amsfonts}
\usepackage{amssymb}
\usepackage{graphicx}
\usepackage{color}
\usepackage{hyperref}
\usepackage{subfigure}

\newcommand{\Tr}{\mbox{Tr}}

\newcommand{\ket}[1]{\left|#1\right\rangle}
\newcommand{\bra}[1]{\left\langle #1\right|}

\begin{document}

\title{A hybrid-systems approach to spin squeezing using a highly dissipative ancillary system}

\author{Shane Dooley}
\email[]{dooleysh@nii.ac.jp}
\affiliation{National Institute of Informatics, 2-1-2 Hitotsubashi, Chiyoda-ku, Tokyo 101-8430, Japan.}

\author{Emi Yukawa}
\affiliation{National Institute of Informatics, 2-1-2 Hitotsubashi, Chiyoda-ku, Tokyo 101-8430, Japan.}

\author{Yuichiro Matsuzaki}
\affiliation{NTT Basic Research Laboratories, 3-1 Morinosato-Wakamiya, Atsugi, Kanagawa 243-0198, Japan.}

\author{George C. Knee}
\affiliation{NTT Basic Research Laboratories, 3-1 Morinosato-Wakamiya, Atsugi, Kanagawa 243-0198, Japan.}

\author{William J. Munro}
\affiliation{NTT Basic Research Laboratories, 3-1 Morinosato-Wakamiya, Atsugi, Kanagawa 243-0198, Japan.}
\affiliation{National Institute of Informatics, 2-1-2 Hitotsubashi, Chiyoda-ku, Tokyo 101-8430, Japan.}

\author{Kae Nemoto}
\affiliation{National Institute of Informatics, 2-1-2 Hitotsubashi, Chiyoda-ku, Tokyo 101-8430, Japan.}

\date{\today}

\begin{abstract}
Squeezed states of spin systems are an important entangled resource for quantum technologies, particularly quantum metrology and sensing. Here we consider the generation of spin squeezed states by interacting the spins with a dissipative ancillary system. We show that spin squeezing can be generated in this model by two different mechanisms: one-axis twisting and driven collective relaxation. We can interpolate between the two mechanisms by simply adjusting the detuning between the dissipative ancillary system and the spin system. Interestingly, we find that for both mechanisms, ancillary system dissipation need not be considered an imperfection in our model, but plays a positive role in spin squeezing. To assess the feasibility of spin squeezing we consider two different implementations with superconducting circuits. We conclude that it is experimentally feasible to generate a squeezed state of hundreds of spins either by one-axis twisting or by driven collective relaxation. 
\end{abstract}


\maketitle

\section{Introduction}
The generation of non-classical states of large quantum systems has attracted significant attention due to the potential of such states in emerging quantum technologies \cite{Dow-03,OBr-09}, such as quantum metrology and sensing \cite{Par-09,Gio-04,Gio-06,Gio-11,Bol-96,Cro-09}. For instance, it is well known that highly entangled states of $N$ spin-$1/2$ particles, such as spin squeezed states, can -- in principle -- be exploited to increase the precision of some measurements by a factor that scales with $N^{1/2}$ compared to the best precision that is achievable with a separable state \cite{Win-92,Win-94,Gio-04,Gio-06,Gio-11}. Interestingly, an improvement in the precision is even possible in the presence of certain types of realistic decoherence, although the scaling of the improvement is reduced to $N^{1/4}$ \cite{Mat-11,Chi-12, Tan-15}. The motivation of the work described here is the generation of such spin squeezed states, starting from an easily prepared separable state of the spin system. 

Solid state spin defects, such as nitrogen vacancy centres or electron donor spins in silicon, are particularly promising candidate spin systems due to their long coherence times \cite{Bar-13,Far-15,Tyr-12}. However, to generate entanglement it is clear that we require some sort of interaction between the spins. Although it has been proposed that this can be achieved using the natural magnetic dipole-dipole interaction between the spins \cite{Cap-09}, in practice this is difficult because any spin will interact very weakly with a distant spin (the strength of the dipole-dipole coupling between two spins scales as $r^{-3}$ where $r$ is the distance between the two spins). Since this interaction is weak it will be challenging to generate highly entangled states within the spin coherence time. Instead, we adopt a hybrid-systems approach where the spins are allowed to interact with an auxilliary system. This interaction with the auxilliary system can induce coupling between the spins (including long range interactions between distant spins) which can then be exploited to generate the necessary entanglement. This approach has been used to experimentally generate few-qubit entangled states of many different systems, for example trapped ions \cite{Lei-05,Mon-11}, Rydberg atoms \cite{Rau-00} and superconducting qubits \cite{Maj-07, Dic-10}. In this context, the auxilliary system is sometimes called a ``quantum bus'' \cite{Spi-06}. However, these experiments are typically limited to few qubit systems. Also, some schemes \cite{Cir-95, Bol-96} need significant entanglement with the ancillary system throughout the interaction. This means that they are limited by the requirement that the ancillary system must have a coherence time that is longer than the duration of the entanglement. In this paper we consider the interaction of a short-lived ancillary system with a long-lived spin system, and we show that this hybrid-systems approach can be used to generate relatively large spin squeezed states. This is a typical feature of the hybrid-systems approach: the strengths of both the auxilliary system and the system of interest are exploited to generate dynamics that would be difficult to generate with either system individially \cite{Xia-13}. 

We structure our paper as follows. In section \ref{sec:bgss} we give our measure of spin squeezing and we introduce the two spin squeezing mechanisms that are relevant to this paper: $(i)$ spin squeezing by one-axis twisting (OAT) \cite{Kit-93,Ma-11}, and $(ii)$ spin squeezing by driven collective relaxation (DCR) \cite{Gon-13,Wol-14}. In section \ref{sec:model} we describe our model and we adiabatically eliminate the ancillary system to obtain an effective master equation for the spin system. We show that both of the spin squeezing mechanisms, OAT and DCR, emerge from these effective dynamics. For concreteness, we focus on two different implementations of the model, one with a superconducting flux qubit playing the role of the ancillary system, and the other with a superconducting microwave resonator. In \ref{sec:OAT} we consider spin squeezing by OAT, including the effect of realistic imperfections in the model, such as dissipation of the ancillary system, inhomogeneity in the spin energies due to fluctuations in their local magnetic fields, and inhomogeneity in the couplings between the ancillary system and the spins. Interestingly, we find that ancillary system dissipation need not be considered an ``imperfection'' in the model. The spin squeezing is very robust to such decoherence and, perhaps counter-intuitively, moderate dissipation can even improve the spin squeezing by OAT. Such ``dissipation-assisted'' spin squeezing is an interesting effect since it is unusual for spin squeezing by OAT to be improved by adding dissipation to a part of the system. We also find that the inhomogeneity in the couplings can be reduced to a negligible level by a judicious experimental setup. Inhomogeneity of the spin energies can be compensated by dynamical decoupling. We find that a pulse sequence known as concatenated-XY8 effectively preserves spin squeezing. However, a drawback of pulsed dynamical decoupling is that, in practice, each pulse in a sequence introduces errors that damage the spin squeezing. At the end of section \ref{sec:OAT} we show that driving the spin system enables spin squeezing by OAT without the need for a dynamical decoupling pulse sequence. In section \ref{sec:DCR} we consider spin squeezing by DCR, including the effect of realistic imperfections in the model. We show that for squeezing by DCR, standard pulse sequences are not effective in overcoming inhomogeneity in the spin energies, but we present a novel pulse sequence that preserves the spin squeezing. We conclude that it is experimentally feasible to generate squeezed states of hundreds of spins, either by OAT or by DCR. Finally -- for our chosen model parameters -- we estimate the improvement in precision that this can give in magnetic field sensing.

\section{Background} \label{sec:bgss}
\begin{figure}[ht]
\vspace{4mm}
\centering
    \includegraphics[width=25mm]{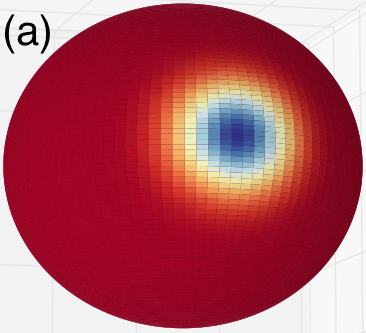}
    \includegraphics[width=25mm]{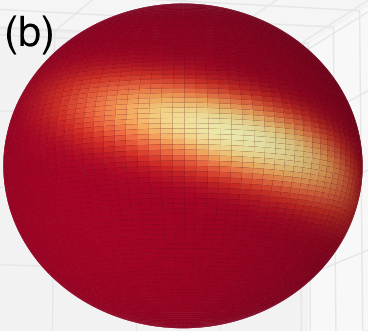}
    \includegraphics[width=25mm]{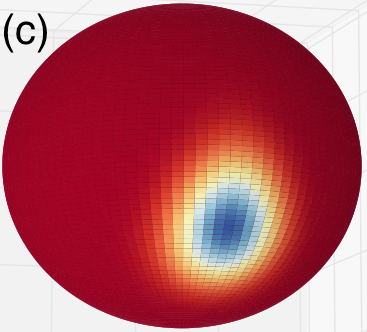}
    \includegraphics[height=23mm]{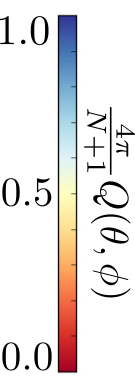}
    \caption{$Q$-functions for $(a)$ a spin coherent state, $(b)$ a one-axis twisted state, and $(c)$ a steady state of driven collective relaxation. The squeezed states have a reduced variance orthogonal to the mean spin direction. The $Q$-function is defined as $Q(\theta,\phi) = \frac{N+1}{4\pi} \bra{\theta,\phi}\rho_s\ket{\theta,\phi}$ for states $\rho_s$ in the $j=N/2$ eigenspace of the operator $\hat{J}^2= \hat{J}_x^2 + \hat{J}_y^2 + \hat{J}_z^2$. For each of these plots $N=40$ and in $(c)$ $\Omega_x = 0$, $2\Omega_y/(\gamma N) = 0.85$.}    
\label{fig:spinsqueezing}
\end{figure}

Our primary system of interest thoughout this paper is an ensemble of $N$ spin-$1/2$ particles. The \emph{collective spin operators} for this system are $\hat{J}_{\mu} = \frac{1}{2}\sum_{i=1}^N \hat{\sigma}_{\mu}^{(i)}$ where $\hat{\sigma}_{\mu}^{(i)}$ are the Pauli operators for the $i$'th spin with $\mu\in\{x,y,z\}$. The \emph{mean spin vector} for an arbitrary state $\rho_s$ of the spin system is the expectation value $\langle\vec{\hat{J}} \rangle = \Tr(\rho_s \vec{\hat{J}})$ where $\vec{\hat{J}} = (\hat{J}_x, \hat{J}_y, \hat{J}_z)$ is a vector of operators. We denote the unit vector in the direction of the mean spin as $\vec{n} = \langle\vec{\hat{J}} \rangle / | \langle\vec{\hat{J}} \rangle |$. We quantify spin squeezing of a state $\rho_s$ with the \emph{Wineland squeezing parameter} \cite{Win-94}, \begin{equation}  \xi^2 = \frac{N}{|\langle \vec{\hat{J}} \rangle|^2} \min_{\vec{n}_{\perp}} \text{Var}(\vec{n}_{\perp} \cdot \vec{\hat{J}}) , \label{eq:W2} \end{equation} where the minimisation is over all unit vectors $\vec{n}_{\perp}$ that are perpendicular to the mean spin direction and $\text{Var}(\vec{n}_{\perp} \cdot \vec{\hat{J}})$ is the variance of the operator $\vec{n}_{\perp} \cdot \vec{\hat{J}}$. For a spin coherent state of the form \cite{Are-72} \begin{equation} \ket{\theta,\phi} = \bigotimes_{i=1}^N \left( \cos\frac{\theta}{2}\ket{\downarrow_i} + e^{-i\phi}\sin\frac{\theta}{2}\ket{\uparrow_i}\right) , \label{eq:scs} \end{equation} where $\ket{\uparrow_i}$ and $\ket{\downarrow_i}$ are the eigenstates of $\hat{\sigma}_z^{(i)}$, we have $|\langle \vec{\hat{J}} \rangle| = N/2$ and $\min_{\vec{n}_{\perp}} \text{Var}(\vec{n}_{\perp} \cdot \vec{\hat{J}}) = N/4$. Hence, by the definition above we have $\xi^2 = 1$ for a spin coherent state. A state is spin squeezed if $\xi^2 < 1$, implying that the variance of the operator $\vec{n}_{\perp}\cdot \vec{\hat{J}}$ (for some choice of $\vec{n}_{\perp}$) is less than that of a spin coherent state. This is illustrated in Fig. \ref{fig:spinsqueezing} where we plot the $Q$-function for a spin coherent state and for two example spin squeezed states. It is also known that the squeezing parameter $\xi^2$ is an entanglement witness, meaning that $\xi^2 < 1$ implies that the (possibly mixed) state $\rho_s$ is entangled. Moreover, the parameter $\xi$ has a specific operational meaning in that it is the ratio of the phase sensitivity of the state $\rho_s$ to that of a spin coherent state in a Ramsey inteferometric measurement \cite{Win-94}. 


There are various possible mechanisms for the generation of spin squeezing starting from a spin coherent state. In this paper we are particularly interested in two of these: $(i)$ spin squeezing by \emph{one-axis twisting} (OAT) \cite{Kit-93,Ma-11} and $(ii)$ spin squeezing by \emph{driven collective relaxation} (DCR) \cite{Gon-13,Wol-14}. 

$(i)$ OAT is generated by the evolution \begin{equation} \dot{\rho}_\text{s} = -\frac{i}{\hbar}[\hat{H}_\text{oat},\rho_\text{s}] , \label{eq:perfectOAT} \end{equation} where the Hamiltonian is quadratic in one of the collective spin operators, for example, $\hat{H}_\text{oat} = \hbar \chi \hat{J}_z^2$. It is well known that for large $N$ this leads to optimum spin squeezing value $\xi^2 \sim N^{-2/3}$ after evolution time \cite{Kit-93, Ma-11}: \begin{equation} t_\text{opt} \approx 3^{1/6}N^{-2/3}/\chi . \label{eq:OATtopt} \end{equation} For the one-axis twisting Hamiltonian $\hat{H}_\text{oat} = \hbar \chi \hat{J}_z^2$ the most spin squeezing is achieved for an initial spin coherent state that is on the equator of the Bloch sphere of each spin, i.e., $\ket{\theta,\phi} = \ket{\frac{\pi}{2}, \phi}$. In Fig. \ref{fig:spinsqueezing}(b) we plot the $Q$-function for a one-axis-twisted state for $N=40$. 

$(ii)$ Spin squeezing by DCR is induced by the Lindblad master equation \begin{equation} \dot{\rho}_s = - i \left[\Omega_x\hat{J}_{x} + \Omega_y \hat{J}_{y}  , \rho_s\right] + \gamma \mathcal{D}[\hat{J}_{-}](\rho_s) , \label{eq:meCR} \end{equation} where $\hat{J}_{\pm} = \sum_{i=1}^N \hat{\sigma}_{\pm}^{(i)}$ are the collective spin raising and lowering operators and the superoperator $\mathcal{D}$ is defined as $\mathcal{D}[\hat{L}](\rho_s) = \hat{L}\rho_s\hat{L}^{\dagger} - \frac{1}{2}\hat{L}^{\dagger}\hat{L}\rho_s - \frac{1}{2}\rho_s\hat{L}^{\dagger}\hat{L}$ for any operator $\hat{L}$. The parameter $\vec{\Omega} = (\Omega_x, \Omega_y, 0)$ is a transverse magnetic field applied to the spin system, and $\gamma$ is the collective spin relaxation rate. Although the model described by Eq. \ref{eq:meCR} has been well-studied \cite{Pur-80,Dru-80,Sch-02}, spin squeezing by this mechanism has been explored only recently \cite{Gon-13, Wol-14}. It has been shown that any initial spin coherent state $\ket{\theta,\phi}$ relaxes to a steady state \cite{Pur-80,Dru-80}. For an appropriate value of the transverse field $|\vec{\Omega}|$, this steady state is squeezed \cite{Gon-13}. In contrast, the analagous driven, dissipative dynamics for a bosonic mode leads to steady states that are always coherent states rather than squeezed states \cite{Ger-05}. In Eq. \ref{eq:meCR}, the steady state with the most squeezing is achieved for $|\vec{\Omega}|$ of the order $N\gamma$ \cite{Gon-13, Wol-14}.  

In the following we see that both the OAT and the DCR spin squeezing mechanisms emerge in the interaction of an ancillary system with a spin ensemble. 

\section{Model}\label{sec:model}
We model the interaction of a spin ensemble with a dissipative ancillary system by the master equation \begin{equation}  \dot{\rho} = -\frac{i}{\hbar} [ \hat{H}, \rho ] + \gamma \mathcal{D}[\hat{A}](\rho) , \label{eq:me1} \end{equation} where $\hat{H} = \hat{H}_\text{s}(t) + \hat{H}_\text{anc} + H_\text{int}$ and: \begin{eqnarray} \hat{H}_\text{s}(t) &=& \frac{\hbar}{2} \sum_{i=1}^N \omega^{(i)} \hat{\sigma}_z^{(i)} + \hbar \Omega \sum_{i=1}^N\hat{\sigma}_x^{(i)}  \cos(\bar{\omega} t + \eta) , \label{eq:Hspins} \\  \hat{H}_\text{anc} &=& \hbar\omega_\text{anc} \hat{A}^{\dagger}\hat{A} , \label{eq:Hanc} \\ \hat{H}_\text{int} &=&  \hbar\sum_{i=1}^N \lambda^{(i)} \, \hat{\sigma}_{x}^{(i)} \otimes ( \hat{A} + \hat{A}^{\dagger} ) . \label{eq:Hint} \end{eqnarray} The operator $\hat{A} = \sum_{n=0}^{d-2}\sqrt{n+1}\ket{n}\bra{n+1}$ is a lowering operator for the $d$-dimensional ancillary system, where $\ket{n}$ is a basis for the ancilla state space. If, for example, the ancillary system is a qubit ($d=2$), the operator $\hat{A}$ is the qubit lowering operator $\hat{\sigma}_{-} = (\hat{\sigma}_x - i\hat{\sigma}_y)/2$, which has the commutation relation $[\hat{\sigma}_{+}, \hat{\sigma}_{-}] = \hat{\sigma}_z$. If the ancillary system is a bosonic mode ($d\to\infty$), the operator $\hat{A}$ is the annihilation operator $\hat{a}$, with the commutation relation $[\hat{a}, \hat{a}^{\dagger}] = 1$. The ancillary system frequency is $\omega_\text{anc}$ and its relaxation rate is $\gamma$ with corresponding relaxation time $\gamma^{-1}$. In Eq. \ref{eq:Hspins} for the spin ensemble, each spin may have a different frequency $\omega^{(i)}$ with an average $\bar{\omega} = \frac{1}{N} \sum_{i=1}^N \omega^{(i)}$ and standard deviation $\delta\omega = \sqrt{\sum_{i=1}^N[\omega^{(i)} - \bar{\omega}]^2/N}$. Also, the spins are driven at the average spin frequency $\bar{\omega}$ by a classical field of amplitude $\Omega$ and phase $\eta$. The interaction Hamiltonian Eq. \ref{eq:Hint} describes the coupling between the spins and the ancillary system, where each spin may have a different coupling $\lambda^{(i)}$, with an average coupling $\bar{\lambda}$ and standard deviation $\delta\lambda$. 

Rotating to an interaction frame defined by the unitary transformation $\hat{U}(t) = \exp\left[ -it \bar{\omega} \left( \frac{1}{2} \sum_{i=1}^N \hat{\sigma}_z^{(i)} + \hat{A}^{\dagger}\hat{A} \right) \right]$, we make a rotating wave approximation which gives the master Eq. \ref{eq:me1} but with the new Hamiltonian: \begin{eqnarray} \hat{H} &=& \hbar\Delta \hat{A}^\dagger \hat{A} +  \hbar\Omega\, \vec{n}_\eta \cdot \vec{\hat{J}}  \nonumber \\ && + \hbar\bar{\lambda} \left(  \hat{J}_{+} \hat{A} +  \hat{J}_{-} \hat{A}^\dagger\right) + \hat{H}_\text{IB} + \hat{H}_\text{IC} , \label{eq:Hrwa} \end{eqnarray} where $\Delta = \omega_\text{anc} - \bar{\omega}$ is the detuning between the ancillary system frequency and the average of the spin frequencies, $\vec{n}_\eta = (\cos\eta,\sin\eta,0)$ is a unit vector in the equatorial plane, and we have defined $\hat{H}_\text{IB} = \frac{\hbar}{2}\sum_{i=1}^N (\omega^{(i)} - \bar{\omega}) \hat{\sigma}_z^{(i)}$ and $\hat{H}_\text{IC} = \hbar \sum_{i=1}^N (\lambda^{(i)} - \bar{\lambda}) \left( \hat{\sigma}_{+}^{(i)} \hat{\sigma}_{-}^\text{FQ} +  \hat{\sigma}_{-}^{(i)} \hat{\sigma}_{+}^\text{FQ} \right)$. In Eq. \ref{eq:Hrwa} the Hamiltonian is separated into an ``inhomogeneous'' part represented by the Hamiltonian terms $\hat{H}_\text{IB}$ and $\hat{H}_\text{IC}$ and a ``homogeneous'' part represented by the remaining terms of Eq. \ref{eq:Hrwa}. The subscript ``IB'' on $\hat{H}_\text{IB}$ stands for ``inhomogeneous broadening''. If each spin has the same frequency $\omega^{(i)} = \bar{\omega}$ (equal to the average value), the inhomogeneous broadening term $\hat{H}_\text{IB}$ vanishes. Similarly, the subscript ``IC'' on $\hat{H}_\text{IC}$ stands for ``inhomogeneous couplings'', and if each spin is equally coupled to the flux qubit we have $\lambda^{(i)} = \bar{\lambda}$ and the inhomogeneous coupling term $\hat{H}_\text{IC}$ vanishes.

We note that spin relaxation and spin dephasing have been neglected in the model described above. If the spins are, for example, an ensemble of donor spins in silicon then this is a reasonable assumption since the spin dephasing time is of the order of seconds and the relaxation time is of the order of tens of minutes at low temperatures \cite{Tyr-12}. Also, long coherence times of $\sim 30 \text{ ms}$ have been achieved for ensembles of nitrogen-vacancy centres in diamond by the use of dynamical decoupling \cite{Far-15}.

\subsection{Effective dynamics}\label{sec:eff_dyn}
To see how spin squeezing is generated in this model, we first define the parameter $\Gamma = \sqrt{\Delta^2 + \gamma^2/4}$. If the ancillary system is initially in its ground state $\ket{0}$ and if $\Gamma$ satisfies the conditions \begin{equation} \Gamma \gg \bar{\lambda} N, \quad  \Gamma \gg \Omega, \quad \Gamma \gg \delta \omega , \label{eq:approxconds} \end{equation} we can adiabatically eliminate the ancillary system. The result is the effective master equation (see Appendix \ref{app:eff_me_deriv} for details) \cite{Aga-97, Rei-12}: \begin{equation} \dot{\rho}_\text{s} = -\frac{i}{\hbar}[\hat{H}_\text{IB} + \hat{H}_\text{eff}, \rho_\text{s}] + \gamma_\text{eff} \mathcal{D}[\hat{J}_{-}](\rho_\text{s}) , \label{eq:effdyn} \end{equation} where $\rho_\text{s} = \Tr_\text{anc}(\rho)$ is the reduced state of the spin system and $\gamma_\text{eff} = \bar{\lambda}^2 \gamma / \Gamma^2$ is the collective-spin relaxation rate. The effective Hamiltonian is \footnote{We note that the effective Hamiltonian Eq. \ref{eq:Heff} bears close similarity to the Lipkin-Meshkov-Glick Hamiltonian \cite{Lip-65}, which has been studied in other contexts \cite{Vid-04, Oru-08}.} \begin{eqnarray} \hat{H}_\text{eff} = \hbar\Omega\, \vec{n}_\eta \cdot \vec{\hat{J}}  + \hbar\chi_\text{eff} \hat{J}_z^2 - \hbar\chi_\text{eff} \hat{J}_z - \hbar \chi_\text{eff} \vec{\hat{J}}\cdot\vec{\hat{J}} , \label{eq:Heff} \end{eqnarray} where $\chi_\text{eff} = \Delta\bar{\lambda}^2/\Gamma^2$. For clarity we have neglected the inhomogeneous couplings, i.e., $\lambda^{(i)} = \bar{\lambda}$ in Eq. \ref{eq:effdyn} and Eq. \ref{eq:Heff}, although the effect of both inhomogeneous couplings and inhomogeneous broadening will be assessed in later numerics.

These effective dynamics have features of both spin squeezing mechanisms that were discussed in section \ref{sec:bgss}, that is, squeezing by OAT and by DCR. The term $\chi_\text{eff}\hat{J}_z^2$ in the effective Hamiltonian is an OAT term, while the collective relaxation in Eq. \ref{eq:effdyn}, in combination with spin drive $\Omega\, \vec{n}_\eta\cdot\vec{\hat{J}}$, are the necessary ingredients for squeezing by DCR (for comparison, see Eq. \ref{eq:meCR}). The two spin squeezing mechanisms appear independently in two different regimes of the effective master Eq. \ref{eq:effdyn}. The OAT regime emerges when the OAT coefficient is much larger than the collective relaxation rate, $\chi_\text{eff} \gg \gamma_\text{eff}$ \cite{Aga-97}. By comparing the expressions for $\chi_\text{eff}$ and $\gamma_\text{eff}$ it is easy to see that this reduces to the condition that the detuning should be much larger than the ancillary system relaxation rate, $\Delta \gg \gamma$. On the other hand, the DCR regime emerges when the collective relaxation dominates the OAT, $\chi_\text{eff} \ll \gamma_\text{eff}$, which corresponds to the condition $\Delta \ll \gamma$. We note that the DCR regime includes, for example, the case where the ancillary system and the spins are resonant ($\Delta = 0$), in which case the effective master Eq. \ref{eq:effdyn} is in the same form as Eq. \ref{eq:meCR}. 


\subsection{Realistic parameters}
To assess the feasibility of spin squeezing we choose some reasonable parameters for our model. We consider two different implementations: $(i)$ a superconducting flux qubit (FQ) coupled to an ensemble of nitrogen-vacancy (NV) centres in diamond, and $(ii)$ a superconducting microwave resonator (MR) coupled to an ensemble of electron donor spins in silicon.

\subsubsection{Superconducting flux qubit and nitrogen-vacancy centres}\label{sec:FQNV}
The spin parameters $\bar{\omega}$ and $\delta\omega$ depend on the type of spin system. Here we take the spins to be NV centres in diamond. Although the NV centre ground state is spin-$1$ \cite{Doh-12}, a magnetic field can be applied to detune one of the spin sub-levels so that -- to a good approximation -- the NV centre can be considered spin-$1/2$. Due to the NV zero-field splitting, we have $\bar{\omega} \approx 2\pi\times 3 \text{ GHz}$ \cite{Doh-13}, and based on recent experimental results we estimate $\delta\omega \approx 2\pi\times3 \text{ kHz}$ \footnote{Recent experiments have measured $\delta\omega = 2\pi\times 200$ kHz for a diamond sample of NV density $\sim 0.67\times 10^{17} \text{ cm}^{-3}$ \cite{Gre-15a}. If the inhomogeneous broadening $\delta\omega$ is due to interaction with substitutional nitrogen atoms (P1 centres) in the diamond lattice (a reasonable assumption for isotopically pure diamond), $\delta\omega$ is approximately linearly related to the NV density \cite{Wyk-97}. For NV density $\sim 10^{15} \text{ cm}^{-3}$ this corresponds to $\delta\omega \approx 2\pi\times 3 \text{ kHz}$.}. The spin drive parameters $\Omega$ and $\eta$ are experimentally tunable. 

For a superconducting FQ, the ancillary system operator $\hat{A}$ in Eqs. \ref{eq:Hanc} and \ref{eq:Hint} is the qubit lowering operator $\hat{\sigma}_{-}$. We assume that the FQ is tuned so that its two persistent current states are degenerate, with tunnel splitting $\omega_\text{anc} = \omega_\text{FQ}$ between them. The detuning $\Delta = \omega_\text{FQ} - \bar{\omega}$ can be varied experimentally by changing the flux qubit tunnel splitting $\omega_\text{FQ}$ \cite{Zhu-10, Fed-10}.

\begin{figure*}[ht]
\centering
    \includegraphics[height=55mm]{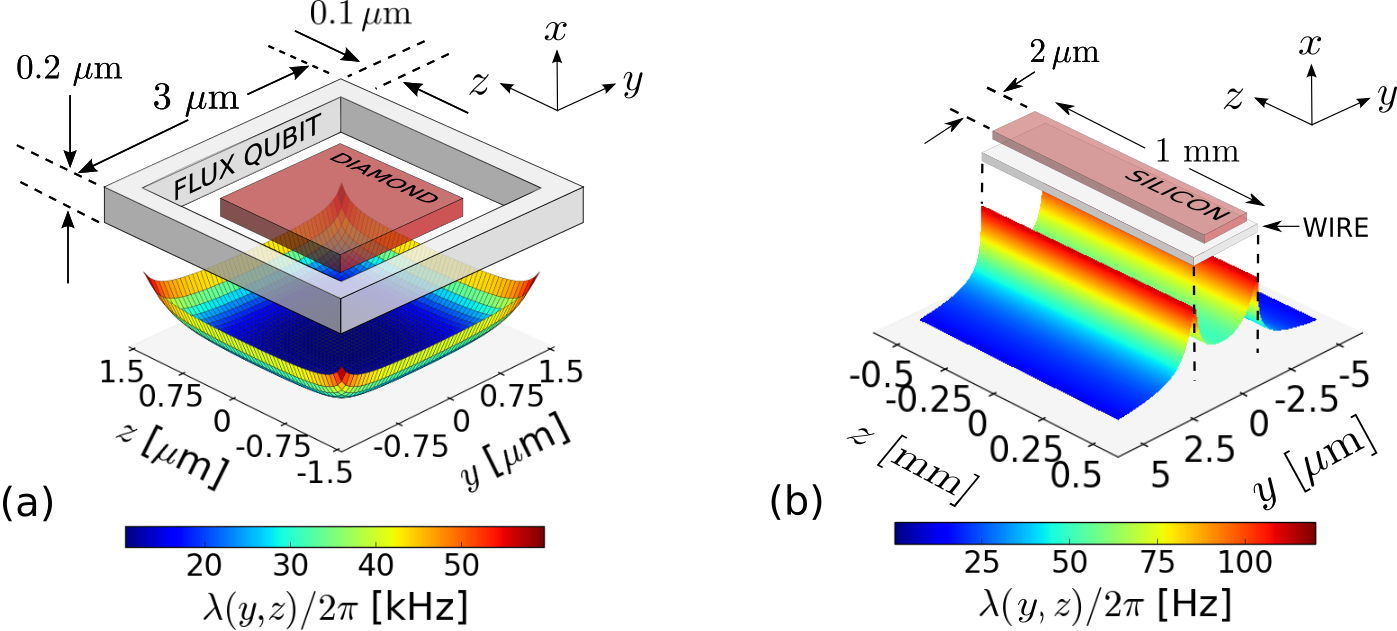}
    \caption{(a) An illustration of the flux qubit model. The flux qubit couples to the NV centres in the diamond sample via the magnetic field generated by the persistent current in the flux qubit. The colourmap shows $\lambda(y,z)/2\pi$, the coupling strength at the coordinates $(0,y,z)$ in the interior of the flux qubit. (b) An illustration of the microwave resonator model. The wire is the central inductance of a superconducting lumped element resonator and generates the magnetic field that couples to the phosphorus electron spins in the silicon crystal. The width, length, and height of the wire are $1.5 \text{ mm}$, $3\, \mu$m, and $50 \text{ nm}$ respectively, and the silicon crystal is positioned $100 \text{ nm}$ above the wire. The colourmap shows $\lambda (y,z) /2\pi$, the coupling strength $100 \text{ nm}$ above the wire.} 
\label{fig:model}
\end{figure*}

The coupling strength $ \lambda^{(i)} = \lambda (y_i,z_i) = g_e \mu_B |B(y_i,z_i)|/(\sqrt{2}\hbar)$ is determined by the magnetic field $B (y_i,z_i)$ that is generated by the FQ at the position $(0,y_i,z_i)$ of the $i$'th NV centre (with axes as shown in Fig. \ref{fig:model}). We assume a square FQ of length $3\,\mu\text{m}$, wire thickness $0.1\,\mu\text{m}$ and wire height $0.2\,\mu\text{m}$ (see figure Fig. \ref{fig:model}(a)) and we assume a uniform critical current of $I = 1.4 \, \mu \text{A}$. Based on these values, the coupling strength $\lambda (y,z)$ in the interior of the FQ can be estimated by the Biot-Savart law \cite{Mar-10}. This is shown in the contour plot in Fig. \ref{fig:model}(a). For NV centres positioned near the middle of the FQ the coupling is relatively homogeneous across a broad area (the blue region in Fig. \ref{fig:model}(a)). Assuming that the NV centres are contained in a diamond sample of volume $1.58 \times 1.58 \times 0.2$ $\mu\text{m}^3$ with NV density $10^{15} \text{ cm}^{-3}$ gives a total of $N = 500$ nitrogen-vacancy centres randomly placed throughout the diamond sample. We find numerically that in this case the average coupling is $\bar{\lambda} \approx 2\pi \times 12$ kHz with standard deviation $\delta \lambda \approx 2\pi \times 1$ kHz. We note that coherent coupling between a FQ and an ensemble of NV centres has been demonstrated experimentally with a similar coupling strength \cite{Zhu-11}.

Finally, we assume that the FQ relaxation rate is $\gamma = 1 \text{ MHz}$, corresponding to the relaxation time $\gamma^{-1} = 1 \, \mu\text{s}$. This is a reasonable estimate, since relaxation times an order of magnitude longer than this have been measured in recent experiments with flux qubits \cite{Byl-11, Ste-14}. We will find it useful to write both the adjustable detuning $\Delta$ and the relaxation rate $\gamma$ as a proportion of the collective coupling $\bar{\lambda}N$. For the relaxation rate this gives $\gamma = 0.0265 \times \bar{\lambda}N$, where $\bar{\lambda} = 2\pi\times 12 \text{ kHz}$ and $N = 500$, as determined above. We use these expressions for $\Delta$ and $\gamma$ when our numerical simulations are limited to small numbers of spins, $N$. This is useful because with these expressions the condition $\Gamma = \sqrt{\Delta^2 + \gamma^2/4} \gg \bar{\lambda}N$ in Eq. \ref{eq:approxconds} is satisfied by the same proportion for any value of $N$, and we can extrapolate from our small-$N$ numerical results to our larger estimated value $N=500$.

\subsubsection{Superconducting microwave resonator and donor spins in silicon}\label{sec:MRdonor}
The coupling of a MR to a spin system is much weaker than the coupling of a FQ to a spin system. For this reason, we take the spin system in this case to be an ensemble of electron spins of phosphorus atoms doped in a silicon crystal, since longer coherence times have been measured for these spins than for NV centres \cite{Tyr-12}. The donor electron spin frequency $\bar{\omega}$ is determined by the electron Zeeman splitting. The donor electron spins interact with the donor nuclear spins via a hyperfine interaction of $2\pi\times 118$ MHz. However, by polarizing the nuclear spins this interaction can be regarded as a contribution to the electron Zeeman splitting \cite{Sim-11}. With an additional externally applied magnetic field $\sim 100 \text{ mT}$ we have $\bar{\omega} \approx 2\pi\times 3 \text{ GHz}$ \cite{Pla-13} (similar to the NV zero-field splitting). Based on experimental results \cite{Tyr-12}, we assume $\delta\omega = 2\pi\times 15 \text{ Hz}$, which is much smaller than for NV centres. The spin drive parameters $\Omega$ and $\eta$ are experimentally tunable.

For a superconducting MR, the operator $\hat{A}$ in Eqs. \ref{eq:Hanc} and \ref{eq:Hint} is the bosonic lowering operator $\hat{a}$. The detuning $\Delta = \omega_\text{MW} - \bar{\omega}$ can be varied experimentally by adjusting the MR frequency $\omega_\text{MW}$. The coupling strength $\lambda^{(i)} = g_e \mu_B |B(y_i,z_i)|/ (2\hbar)$ is determined by the magnetic field $B(y_i,z_i)$ that is generated by the MR at the position $(0,y_i,z_i)$ of the $i$'th donor spin. We assume that the wire of the MR has length $1.5 \text{ mm}$, width $2\,\mu$m, height $50 \text{ nm}$ (see Fig. \ref{fig:model}(b)), a penetration depth of $90 \text{ nm}$, and an inductance $L = 1.5 \text{ nH}$. Based on these values, the coupling strength is shown in the contour plot in Fig. \ref{fig:model}(b). There is a region of relatively homogeneous coupling directly above the wire (the green area between the two ridges in the contour plot). We suppose that a silicon sample of dimensions $1 \text{ mm} \times 2\,\mu\text{m} \times 50 \text{ nm}$ and donor spin density $1.2 \times 10^{14} \text{ cm}^{-3}$ is placed in this region at a distance of $100 \text{ nm}$ from the resonator. With these values we estimate that there are $N = 1.2\times 10^4$ spins placed randomly throughout the silicon sample, with an average coupling $\bar{\lambda} \approx 2\pi \times 56 \text{ Hz}$ and a standard deviation $\delta\lambda \approx 2\pi \times 4 \text{ Hz}$ \cite{Bie-15}. This is a considerably weaker coupling than for the FQ and NV implementation in the previous section. However, measured coherence times for donor spins in silicon are much longer than those for NV centres \cite{Tyr-12}. The relaxation rate of the resonator is $\gamma = \omega_\text{MR} / Q \approx 2 \pi \times 0.34 \text{ MHz}$, assuming a resonator quality factor $Q = 4.5 \times 10^4$ and the frequency $\omega_\text{MR} \approx 2\pi \times 3 \text{ GHz}$. For $\bar{\lambda} = 2\pi\times 56 \text{ Hz}$ and $N= 1.2 \times 10^4$ this relaxation rate can be expressed as $\gamma = 0.1 \bar{\lambda}N$. We use this value when our numerical simulations are restricted to small values of $N$.

In the following sections we numerically investigate the spin squeezing, using realistic parameters as far as possible. For all of our simulations we use the master Eq. \ref{eq:me1} with the Hamiltonian Eq. \ref{eq:Hrwa}, that is, the master equation \emph{before} the approximations that leads to the effective master Eq. \ref{eq:effdyn}. This gives us meaningful results even when the approximation conditions in Eq. \ref{eq:approxconds} are not well satisfied.

\begin{figure*}[!ht]
\centering
    \includegraphics[height=53mm]{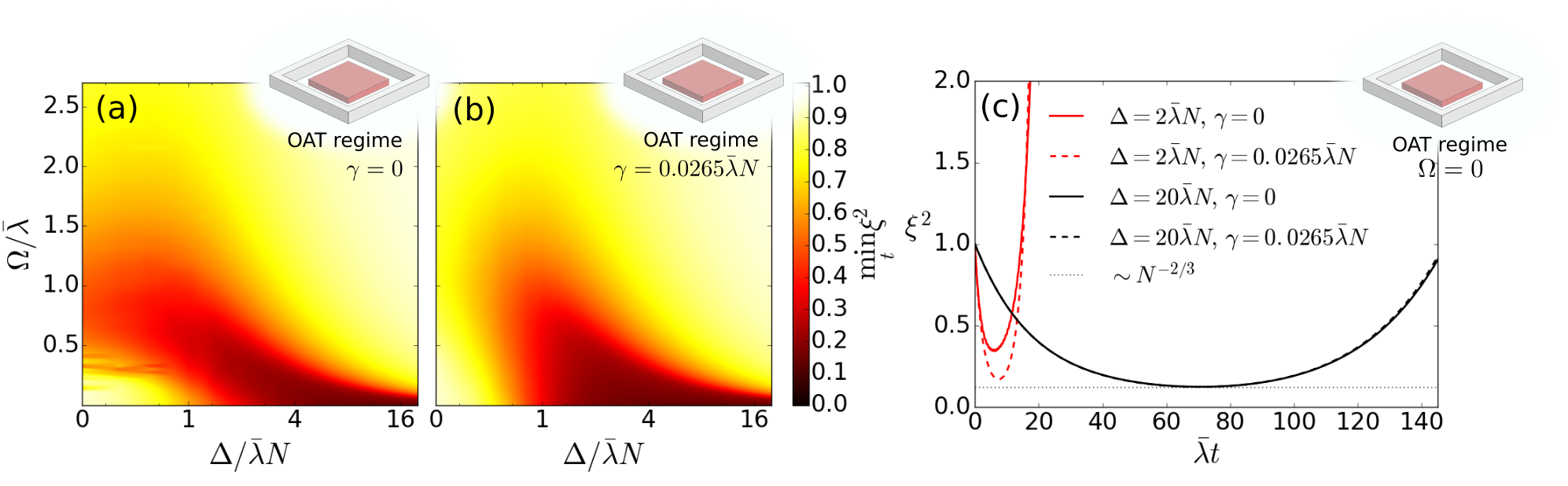}
    \caption{Spin squeezing in the OAT regime for initial spin coherent state $\ket{\theta, \phi} = \ket{\frac{\pi}{2}, 0}$. (a), (b) The spin squeezing that can be achieved for various values of the detuning $\Delta$ and drive $\Omega$ for ancillary system relaxation (a) $\gamma = 0$, (b) $\gamma = 0.0265\times\bar{\lambda}N$. (c) The time evolution of the spin squeezing parameter for $\Omega = 0$ and for two different values of $\Delta$. Comparing the solid lines with the dashed lines we see that the spin squeezing is robust to ancillary system relaxation and can even be improved by it (solid red line vs. dashed red line). All figures were plotted using master Eq. \ref{eq:me1} and Hamiltonian Eq. \ref{eq:Hrwa} and for a flux qubit ancillary system with $N=40$, $\bar{\lambda} = 2\pi\times 12$ kHz, $\eta = 0$, $\delta\lambda = 0$, $\delta\omega = 0$. For comparison, the horizontal black dotted line in (c) shows the level of optimum spin squeezing for perfect OAT, i.e., the optimum squeezing due to the effective Hamiltonian Eq. \ref{eq:Heff}.}    
\label{fig:OAT_limit_plus}
\end{figure*}

\section{Spin squeezing by OAT} \label{sec:OAT}
\subsection{Ideal case, with ancilla relaxation}\label{sec:OAT_plus_ideal}
First, we consider spin squeezing in the OAT regime $\Gamma \approx \Delta \gg \gamma$, assuming that the initial state is the spin coherent state $\ket{\theta, \phi} = \ket{\frac{\pi}{2},0}$, since this state is in the class of states $\ket{\frac{\pi}{2},\phi}$ that leads to the most squeezing by the OAT term $\chi_\text{eff}\hat{J}_z^2$. To prepare the state $\ket{\frac{\pi}{2},0}$ we make use of the fact that a general spin coherent state $\ket{\theta,\phi} = \hat{R}(\theta,\phi)\ket{0,0}$ can be prepared by rotating each spin from the state $\ket{0,0} = \bigotimes_{i=1}^N \ket{\downarrow_i}$ with the rotation operator $\hat{R}(\theta,\phi) = \exp[-i\theta(\hat{J}_x\sin\phi - \hat{J}_y\cos\phi)]$ \cite{Are-72}. This rotation can be implemented by applying an electomagnetic pulse to the spin system. The state $\ket{\theta, \phi} = \ket{0, 0}$ is itself easily prepared, e.g., by cooling (it is the ground state of the spin Hamiltonian Eq. \ref{eq:Hspins} when $\Omega = 0$) or, for NV centres, by optical pumping \cite{Rob-11,Doh-13}. After the state $\ket{\frac{\pi}{2},0} = \hat{R}(\frac{\pi}{2},0)\ket{0,0}$ has been prepared, unitary evolution by $\hat{H}_\text{eff}$ leads to spin squeezing.

As mentioned in the previous section, the detuning $\Delta$ and the spin drive $\Omega$ are experimentally tunable parameters in both of the considered implementations. To get a comprehensive picture of which values of these parameters lead to spin squeezing we plot $\min_{t}\xi^2$, the minimum spin squeezing that is achieved across all evolution times $t$, as a function of $\Delta$ and $\Omega$. This is shown in Fig. \ref{fig:OAT_limit_plus}(a) for the FQ and NV model assuming (for the moment) the ideal case where there is no flux qubit relaxation ($\gamma = 0$), no inhomogeneous broadening ($\delta\omega = 0$) and homogeneous coupling of the spins to the flux qubit ($\delta\lambda = 0$). Fig. \ref{fig:OAT_limit_plus}(a) shows that there is significant spin squeezing (the dark red region) in the lower right portion of the plot where $\Gamma \approx \Delta  \gg \Omega $ and $\Gamma \approx \Delta \gg \bar{\lambda} N$, corresponding the the regime of validity of the effective Hamiltonian Eq. \ref{eq:Heff}. In Fig. \ref{fig:OAT_limit_plus}(b) we include a realistic amount of flux qubit relaxation ($\gamma = 0.0265 \times \bar{\lambda} N$) and we see that the spin squeezing is very robust to this kind of decoherence. The corresponding plots for the resonator model are not shown as they are qualitatively similar to Figs. \ref{fig:OAT_limit_plus}(a,b).

In the OAT regime ($\Delta\gg\gamma$) the evolution time required to reach the optimal spin squeezing is given by Eq. \ref{eq:OATtopt}. From this equation we estimate \begin{equation} t_\text{opt} = 3^{1/6} N^{-2/3}\Delta/\bar{\lambda}^2 , \label{eq:topt2} \end{equation} where we have substituted the expression for the OAT coefficient $\chi_\text{eff}$ and we have used $\Gamma\approx\Delta$. It may appear from this expression for $t_\text{opt}$ that the optimum squeezing time decreases with the number of spins $N$, but this is not so, since we require $\Gamma\approx\Delta \gg \bar{\lambda} N$ for the effective Hamiltonian Eq. \ref{eq:Heff} to be valid. For a detuning $\Delta = k \bar{\lambda} N \gg \bar{\lambda} N$ for some $k\gg 1$, this translates to an optimal squeezing time $t_\text{opt} = 3^{1/6} k N^{1/3}/\bar{\lambda}$. We see that the optimum squeezing time actually increases as the number of spins $N$ increases. However, the scaling is $N^{1/3}$ so that $t_\text{opt}$ is not too large for moderate values of $N$. It was determined above that $N=500$ was a realistic number of NV centres that could be coupled to the FQ for our chosen FQ dimensions. In this case, substituting $\Delta = 20\bar{\lambda}N$ and $\bar{\lambda} = 2\pi\times 12 \text{ kHz}$, we estimate $t_\text{opt} = 2.5 \text{ ms}$, which is within the spin coherence times achieved in recent experiments with ensembles of nitrogen-vacancy centres in diamond \cite{Far-15}. In Fig. \ref{fig:OAT_limit_plus}(c) the solid black line shows the time evolution of the spin squeezing parameter, assuming the large detuning $\Delta = 20\bar{\lambda}N$, zero ancillary system relaxation $\gamma = 0$, and zero spin drive $\Omega = 0$. For comparison, the dashed black line shows the spin squeezing for relaxation $\gamma = 0.0265 \times \bar{\lambda} N$. The spin squeezing is almost indistinguishable from the $\gamma = 0$ case, confirming that this spin squeezing mechanism is very robust to realistic levels of ancillary system relaxation. Also, the horizontal dotted black line shows the level of optimum spin squeezing $\xi^2 \sim N^{-2/3}$ for perfect OAT, i.e., the optimum squeezing by the effective Hamiltonian Eq. \ref{eq:Heff} (or, alternatively, by Eq. \ref{eq:perfectOAT}). The minimum of the solid black line reaches this optimum since the dynamics are well approximated by the effective Hamiltonian for the large detuning $\Delta = 20\bar{\lambda}N$. For the microwave resonator and donor spins in silicon implementation, we estimated $N = 1.2\times 10^4$ and $\bar{\lambda} = 2\pi\times 56 \text{ Hz}$. Substituting these values, along with $\Delta = 20\bar{\lambda}N$, gives $t_\text{opt} = 1.6 \text{ s}$, which is within the spin coherence time $\sim 10 \text{ s}$ of phosphorus donor spins in silicon at low temperatures with spin echo \cite{Tyr-12}. 

Although these optimum squeezing times are within the achievable spin coherence times in both implementations, it is desirable to decrease $t_\text{opt}$. Interestingly, we observe that for a given number of spins we can significantly reduce the optimum squeezing time as follows. Since the optimum squeezing time $t_\text{opt}$ scales linearly with the detuning $\Delta$ (see Eq. \ref{eq:topt2}), the spins can be squeezed more quickly by decreasing the detuning. Decreasing the detuning comes at a cost, however: we need $\Delta$ to be large enough to satisfy our approximation condition $\Gamma \approx \Delta \gg \bar{\lambda} N$. This leads to a tradeoff: if we decrease the detuning we can squeeze more quickly at the expense of a worse approximation to the effective Hamiltonian, and conversely, if we increase the detuning we have a better approximation but at the expense of a longer wait for the optimal squeezing. This can be seen by comparing the solid black line and the solid red line in Fig. \ref{fig:OAT_limit_plus}(c) for the FQ and NV model. The detuning $\Delta = 2 \bar{\lambda} N$ for the solid red line, is an order of magnitude smaller than $\Delta = 20 \bar{\lambda} N$ for the solid black line, so that the optimum spin squeezing is achieved an order of magnitude faster. However, the minimum spin squeezing $\min_t\xi^2$ is degraded compared to the optimal squeezing (the horizontal black dotted line). This difference between the optimum and the minumum of the solid red curve shows the importance of using the full master Eq. \ref{eq:me1} instead of the effective master Eq. \ref{eq:effdyn} for smaller values of $\Delta$. Interestingly, if $\Delta$ is not quite large enough to satisfy $\Delta \gg\bar{\lambda} N$, flux qubit relaxation can significantly improve the approximation so long as the effect of collective relaxation on the OAT is still negligible, i.e, provided that $\Delta \gg \gamma$. This is shown by the dotted red line in Fig. \ref{fig:OAT_limit_plus}(c) for the FQ and NV model with $\gamma = 0.0265 \times \bar{\lambda} N$. With this realistic amount of flux qubit relaxation the squeezing can be significantly improved compared to $\gamma = 0$ (the solid red line). This improvement of the spin squeezing by ancillary system relaxation may be surprising on first sight, since in most models any form of decoherence is an unwanted influence on the dynamics. However, the effect of the relaxation is to suppress excitation of the ancillary system. This, in turn, inhibits entanglement between the ancillary system and the spin system, which would be damaging to the spin squeezing. If we choose $\Delta = 2\bar{\lambda}N$ the optimum squeezing time for $N=500$ spins with the FQ and NV implementation is reduced to $t_\text{opt} = 250\, \mu\text{s}$. Similarly, the optimum squeezing time for the MR and donor spins implementation is decreased to $t_\text{opt} = 160 \text{ ms}$. 

\subsection{Realistic case, with dynamical decoupling} \label{sec:OAT_dd}
\begin{figure}[]
\centering
	\includegraphics[height=50mm]{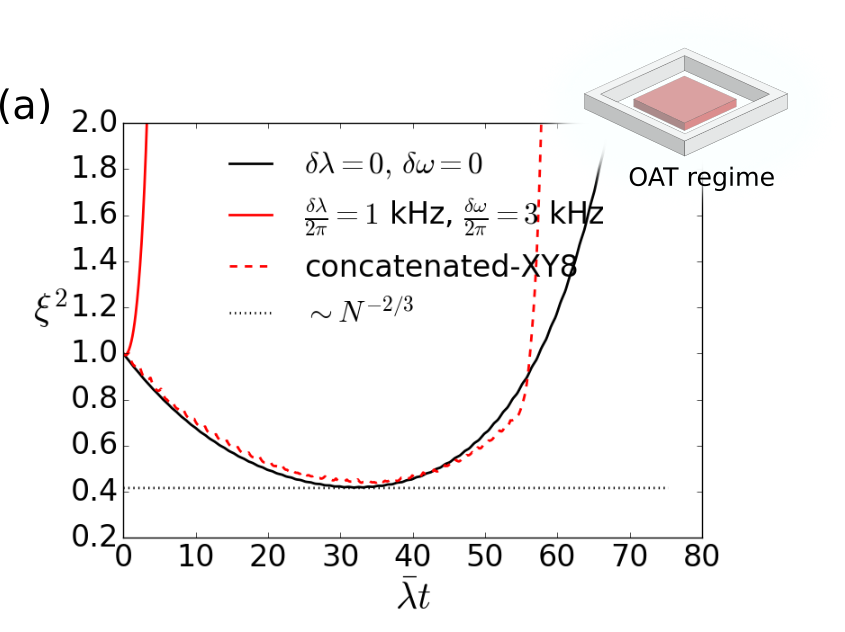}
        \includegraphics[height=50mm]{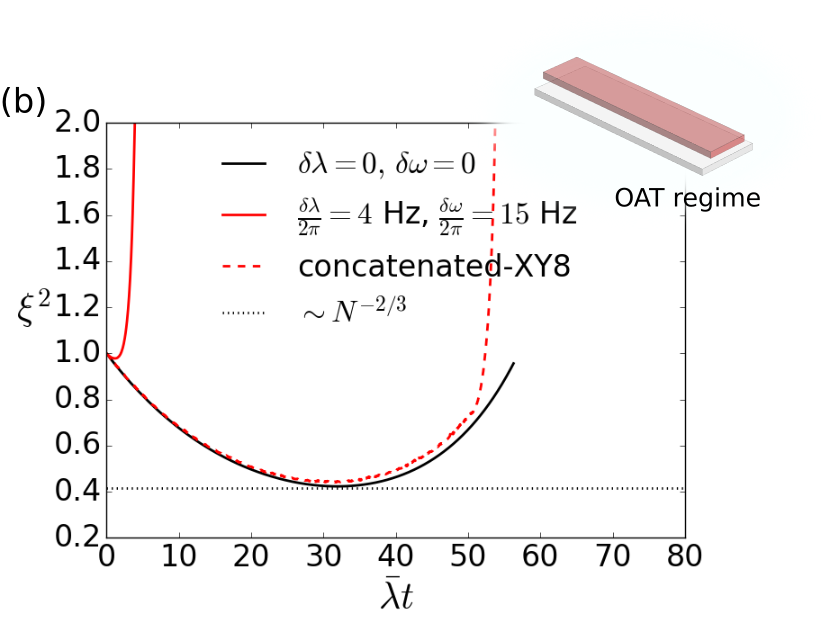}
        \caption[]{For both (a) the FQ and NV implementation, and (b) the MR and donor spins implementation, the spin squeezing is badly degraded by inhomogeneous broadening (the solid red lines) but can be completely recovered by dynamical decoupling with a concatenated-XY8 pulse sequence (dashed red lines). The XY8 pulse sequence is $C = \hat{\pi}_x \text{--} \hat{\pi}_y \text{--} \hat{\pi}_x \text{--} \hat{\pi}_y \text{--} \hat{\pi}_y \text{--} \hat{\pi}_x \text{--} \hat{\pi}_y \text{--} \hat{\pi}_x$, where $\hat{\pi}_x$ represents the $\pi$-pulse $\hat{R}(\pi,\pi/2) = \exp[-i\pi\hat{J}_x]$, $\hat{\pi}_y$ represents the $\pi$-pulse $\hat{R}(\pi,0) = \exp[i\pi\hat{J}_y]$, and each dash represents free evolution for a time $\tau$. The concatenated-XY8 pulse sequence is \cite{Far-15} $C \text{--} \hat{\pi}_x \text{--} C \text{--} \hat{\pi}_y \text{--} C \text{--} \hat{\pi}_x \text{--} C \text{--} \hat{\pi}_y \text{--} C \text{--} \hat{\pi}_y \text{--}C \text{--} \hat{\pi}_x \text{--}C \text{--} \hat{\pi}_y \text{--}C \text{--} \hat{\pi}_x $, a sequence of 72 $\pi$-pulses in total, ending at $\bar{\lambda}t \approx 50$ in both (a) and (b). Each line in the figures is averaged over $100$ evolutions to remove random fluctuations. Horizontal dotted lines show the optimum spin squeezing by OAT. (Fig. (a) parameters: $N=6$, $\bar{\lambda}=2\pi\times 12$ kHz, $\gamma = 0.0265\times\bar{\lambda}N$, $\Delta = 20 \bar{\lambda}N$, $\Omega = 0$, $\eta = 0$, $\tau = 0.01 \text{ ms}$; Fig. (b) parameters: $N = 6$, $\bar{\lambda}=2\pi\times 56$ Hz, $\gamma = 0.1\times\bar{\lambda}N$, $\Delta = 20 \bar{\lambda}N$, $\Omega = 0$, $\eta = 0$, $\tau = 1 \text{ ms}$.)}
\label{fig:dd}
\end{figure}

We now consider the effect of inhomogeneous broadening and inhomogeneous couplings on the OAT spin squeezing. In this case, since the state space dimension increases exponentially in the number of spins, $N$, our numerics are restricted to a small number of spins, $N=6$. In Fig. \ref{fig:dd}(a) the solid red line shows the spin squeezing including the effect of inhomogeneous couplings $\hat{H}_\text{IC}$ and inhomogeneous broadening $\hat{H}_\text{IB}$ in the FQ and NV model, using the value $\delta\lambda = 2\pi\times 1 \text{ kHz}$ that was estimated for the standard deviation of the couplings $\lambda^{(i)}$, and spin frequencies $\omega_i$ chosen at random from a Gaussian distribution with standard deviation $\delta\omega = 2\pi\times 3 \text{ kHz}$. The dynamics are averaged over 100 evolutions to remove fluctuations due to the randomness of the $\omega_i$. Similarly, the solid red line in Fig. \ref{fig:dd}(b) shows the spin squeezing including inhomogeneities with the MR model parameters. For both implementations, we see that the spin squeezing is degraded, although there is a small amount of spin squeezing at very short times for the MR model parameters in Fig. \ref{fig:dd}(b). Further numerics have shown that the decay of spin squeezing is primarily due to the inhomogeneous broadening term $\hat{H}_\text{IB}$ rather than inhomogeneous coupling term $\hat{H}_\text{IC}$. In fact, for the relatively homogeneous couplings achieved in the setups illustrated in Fig. \ref{fig:model}, the inhomogeneous coupling term $\hat{H}_\text{IC}$ can be safely neglected for timescales of interest. To understand the damaging effect of the inhomogeneous broadening we note that the Hamiltonian $\hat{H}_\text{IB}$ causes each spin to evolve around its Bloch sphere at a different rate determined by the frequency $\omega_i - \bar{\omega}$. For a Gaussian distibution of $\omega_i$, this dephasing leads to Gaussian decay of $|\langle\vec{\hat{J}}\rangle|$, the magnitude of the mean spin vector, with a decay time $(\delta\omega)^{-1}$. Since the Wineland squeezing parameter $\xi^2$ is inversely proportional to $|\langle\vec{\hat{J}}\rangle|^2$, decay of the mean spin vector leads to an increase in the squeezing parameter.

It is well known that for evolution by $\hat{H}_\text{IB}$, a single ${\pi}$-pulse at some time $\tau$ leads to a spin echo at the time $2\tau$, where the ${\pi}$-pulse is an instantaneous rotation $\hat{R}(\pi,\phi) = \exp[-i\pi (\hat{J}_x\sin\phi - \hat{J}_y\cos\phi)]$ by an angle $\pi$ about an axis on the equator of the Bloch sphere of each spin. Crucially, the ${\pi}$-pulse also commutes with the OAT operator $\hat{J}_z^2$ so that a ${\pi}$-pulse at some time $\tau$ has the effect of undoing the inhomogeneous broadening at time $2\tau$ without affecting the spin squeezing by OAT \cite{Ben-13}. To see this we consider the effective master Eq. \ref{eq:effdyn} in the OAT regime $\Delta \gg \gamma$. Assuming $\Omega = 0$, $\delta\lambda = 0$ and neglecting collective relaxation, the spin system evolves by \begin{equation} \dot{\rho}_\text{s} = -\frac{i}{\hbar}[\hat{H}_\text{IB} + \hbar\chi_\text{eff}\hat{J}_z^2 - \hbar\chi_\text{eff}\hat{J}_z - \hbar\chi_\text{eff}\vec{\hat{J}}\cdot\vec{\hat{J}} , \rho_\text{s}] . \label{eq:OAT_t<tau} \end{equation} If at time $\tau$ we apply the $\pi$-pulse operator $\hat{R}(\pi,\phi)$, the state is transformed to $\rho'_\text{s}(\tau) = \hat{R}(\pi,\phi)\rho_\text{s}(\tau)\hat{R}^\dagger(\pi,\phi)$ and the evolution equation for the following period of time $t>\tau$ is: \begin{equation} \dot{\rho}'_\text{s} = -\frac{i}{\hbar}[\hat{H}_\text{IB} + \hbar\chi_\text{eff}\hat{J}_z^2 - \hbar\chi_\text{eff}\hat{J}_z - \hbar\chi_\text{eff}\vec{\hat{J}}\cdot\vec{\hat{J}} , \rho'_\text{s}] . \label{eq:OAT_t>tau_1} \end{equation} Operating on Eq. \ref{eq:OAT_t>tau_1} on the left by $\hat{R}^\dagger(\pi,\phi)$ and on the right by $\hat{R}(\pi,\phi)$ gives, for $t > \tau$, the evolution equation: \begin{equation}  \dot{\rho}_\text{s} = -\frac{i}{\hbar}[- \hat{H}_\text{IB} + \hbar\chi_\text{eff}\hat{J}_z^2 + \hbar\chi_\text{eff}\hat{J}_z - \hbar\chi_\text{eff}\vec{\hat{J}}\cdot\vec{\hat{J}} , \rho_\text{s}] , \label{eq:OAT_t>tau_2}\end{equation} where we have used $\hat{R}^\dagger(\pi,\phi)\hat{H}_\text{IB} \hat{R}(\pi,\phi) = - \hat{H}_\text{IB}$, $\hat{R}^\dagger(\pi,\phi)\hat{J}_z \hat{R}(\pi,\phi) = - \hat{J}_z$, $\hat{R}^\dagger(\pi,\phi)(\vec{\hat{J}}\cdot\vec{\hat{J}})\hat{R}(\pi,\phi)= \vec{\hat{J}}\cdot\vec{\hat{J}}$, and the important property $\hat{R}^\dagger(\pi,\phi)\hat{J}_z^2 \hat{R}(\pi,\phi) = \hat{J}_z^2$. Comparing Eq. \ref{eq:OAT_t<tau} and Eq. \ref{eq:OAT_t>tau_2} shows that the effect of the $\pi$-pulse is to reverse the sign of the inhomogeneous broadening Hamiltonian $\hat{H}_\text{IB}$ in the following period of evolution, without changing the OAT operator $\hat{J}_z^2$. Eqs. \ref{eq:OAT_t<tau} and \ref{eq:OAT_t>tau_2} are easily solved to give the combined unitary evolution operator: \begin{eqnarray} && \hat{U}(t) = \nonumber\\ && \exp\left[-\frac{i(t - \tau)}{\hbar}(- \hat{H}_\text{IB} + \hbar\chi_\text{eff}\hat{J}_z^2 + \hbar\chi_\text{eff}\hat{J}_z - \hbar\chi_\text{eff}\vec{\hat{J}}\cdot\vec{\hat{J}})\right] \times \nonumber\\ && \exp\left[-\frac{i\tau}{\hbar}(\hat{H}_\text{IB} + \hbar\chi_\text{eff}\hat{J}_z^2 - \hbar\chi_\text{eff}\hat{J}_z - \hbar\chi_\text{eff}\vec{\hat{J}}\cdot\vec{\hat{J}})\right] , \nonumber \end{eqnarray} for times $t>\tau$. The operators in the two exponents above commute, so that at $t=2\tau$ we have \begin{equation} \hat{U}(2\tau) = \exp\left[-\frac{i2\tau}{\hbar} (\hbar\chi_\text{eff}\hat{J}_z^2 - \hbar\chi_\text{eff}\vec{\hat{J}}\cdot\vec{\hat{J}}) \right] . \end{equation} We see that at this time there is a spin-echo (the inhomogeneous broadening Hamiltonian $\hat{H}_\text{IB}$ has been cancelled), but that the OAT squeezing is unaffected. However, in a real system the higher order terms in the effective Hamiltonian will also contribute to the dynamics. We have performed numerics that show that for evolution by the Hamiltonian Eq. \ref{eq:Hrwa}, which includes higher-order terms, a single ${\pi}$-pulse does not completely recover the spin squeezing even if the approximation conditions Eq. \ref{eq:approxconds} are satisfied. This is because the single ${\pi}$-pulse does not refocus the spin dephasing due to higher order inhomogeneous terms in the effective dynamics. To fully preserve the spin squeezing a more complicated dynamical decoupling pulse sequence is required. We have tried various pulse sequences numerically and found that spin squeezing can be completely recovered for a sequence of alternating $\hat{R}(\pi,0)$ and $\hat{R}(\pi,\pi/2)$ pulses, as shown in Fig. \ref{fig:dd} (the dashed red line). The pulse sequence, known as concatenated-XY8, has recently been implemented experimentally to increase the coherence time of an ensemble of nitrogen-vacancy centres to $\sim 30 \text{ ms}$ \cite{Far-15}.

Finally, we note that this model for OAT has been discussed by previous authors for an ancillary bosonic mode \cite{Aga-97, Ben-13}. However, compared to previous work, we highlight the robustness of spin squeezing to ancillary system dissipation, and we demonstrate that ancillary system dissipation can even play a positive role in the generation of spin squeezed states. We have also demonstrated the feasibility of spin squeezing in the two implementations considered, and we have shown that inhomogeneous broadening can be overcome by the concatenated-XY8 pulse sequence.

\begin{figure*}[!ht]
\centering
    \includegraphics[height=50mm]{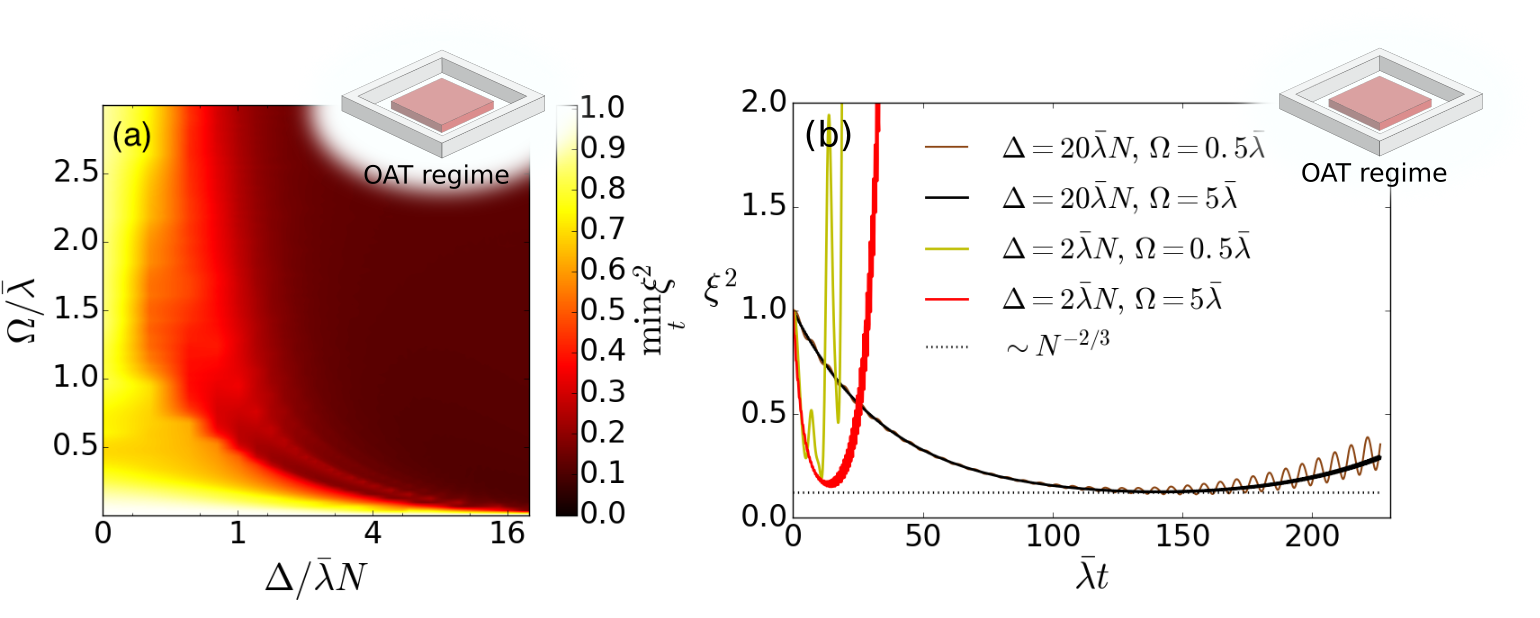}
    \caption{Spin squeezing in the OAT regime ($\Delta\gg\gamma$) for initial spin coherent state $\ket{\theta, \phi} = \ket{0, 0}$. (a) The squeezing that can be achieved for various values of the detuning $\Delta$ and drive $\Omega$. (b) The time evolution of the squeezing parameter for several values of $\Delta$ and $\Omega$. Both (a) and (b) are plotted with homogeneous couplings ($\delta\lambda = 0$), no inhomogeneous broadening ($\delta\omega = 0$), and $N=40$. Both figures were plotted for a flux qubit ancillary system with $\bar{\lambda} =2\pi\times 12$ kHz and $\gamma = 0.0265\times\bar{\lambda}N$, using master Eq. \ref{eq:me1} and Hamiltonian Eq. \ref{eq:Hrwa}. For comparison, the horizontal dotted black line in (b) shows the optimum spin squeezing by OAT.}    
\label{fig:OAT_limit_ground}
\end{figure*} 

\subsection{Realistic case, without dynamical decoupling pulses}\label{sec:OAT_ground}
A practical challenge in the spin squeezing method outlined above is the application of accurate dynamical decoupling pulses to the spin system. For example, we have assumed that each $\pi$-pulse has no errors and that it can be implemented instantaneously. In reality, however, a $\pi$-pulse cannot be implemented instantaneously, and if there are errors in each of the pulses in a sequence, these errors may accumulate, leading to degradation of the spin squeezing. Moreover, the preparation of the spin coherent state $\ket{\theta,\phi}=\ket{\frac{\pi}{2},0}$ requires a pulse that rotates each of the spins equally. If each spin is rotated by a slightly different angle, this introduces inhomogeneous broadening to the system and damages the spin squeezing. In this section we suggest an alternative approach that generates spin squeezing by OAT without the need for dynamical decoupling pulses and starting from the spin coherent state $\ket{\theta,\phi} = \ket{0,0}$ that can be prepared without applying a pulse to rotate the spins.

To show how this works, we derive another effective Hamiltonian in the OAT regime ($\Gamma \approx \Delta \gg \gamma$) starting from $\hat{H}_\text{eff}$ (Eq. \ref{eq:Heff}). First, we rotate $\hat{H}_\text{eff}$ to an interaction frame defined by the unitary transformation $\hat{U}'(t) = \exp [ -it \Omega\, \vec{n}_\eta \cdot \vec{\hat{J}}]$. In the rotating frame the Hamiltonian is \begin{eqnarray} \hat{H}'_\text{eff}  &=& \hbar \chi_\text{eff} \hat{U}^{\prime\dagger}(t) \left( \hat{J}_z^2 - \hat{J}_z -  \vec{\hat{J}}\cdot\vec{\hat{J}} \right) \hat{U}'(t) \\ &=& \hbar\chi_\text{eff} \Big\{ -\frac{1}{2}(\vec{n}_{\eta}\cdot \vec{\hat{J}})^2 -  \frac{1}{2} \vec{\hat{J}}\cdot\vec{\hat{J}} - \cos(\Omega t)\hat{J}_z \nonumber\\ &&  - \sin(\Omega t) (\vec{n}_{\eta + \frac{\pi}{2}} \cdot \vec{\hat{J}}) + \frac{\cos(2\Omega t)}{2} [\hat{J}_z^2  - (\vec{n}_{\eta + \frac{\pi}{2}}\cdot\vec{\hat{J}})^2] \nonumber\\ &&  + \frac{\sin(2\Omega t)}{2} [ \hat{J}_z (\vec{n}_{\eta + \frac{\pi}{2}} \cdot \vec{\hat{J}}) + (\vec{n}_{\eta + \frac{\pi}{2}} \cdot \vec{\hat{J}})\hat{J}_z  ] \Big\} \label{eq:Heff2a} \end{eqnarray} If the parameter $\Omega$ is large enough to satisfy the condition $\Omega \gg N \chi_\text{eff}$ we can make a rotating wave approximation by neglecting quickly oscillating terms in Eq. \ref{eq:Heff2a}. The resulting effective Hamiltonian is:  \begin{equation} \hat{H}'_{\text{eff}} \approx - \frac{\hbar\chi_\text{eff}}{2}(\vec{n}_\eta \cdot \vec{\hat{J}})^2 - \frac{\hbar\chi_\text{eff}}{2} \vec{\hat{J}} \cdot \vec{\hat{J}} . \label{eq:Heff2} \end{equation} Intuitively, this is the effective Hamiltonian that results from averaging $\hat{H}_\text{eff}$ over rapid oscillations around the $\vec{n}_\eta$-axis due to the large drive $\Omega$. Compared to the effective Hamiltonian $\hat{H}_\text{eff}$ (Eq. \ref{eq:Heff}), the key feature of $\hat{H}'_{\text{eff}}$ (Eq. \ref{eq:Heff2}) is that the OAT term is $(\vec{n}_\eta \cdot \vec{\hat{J}})^2$ rather than $\hat{J}_z^2$. For instance, if $\eta = 0$ we have $(\vec{n}_{0} \cdot \vec{\hat{J}})^2 = \hat{J}_{x}^2$, or if $\eta = \pi/2$ we have $(\vec{n}_{\pi/2} \cdot \vec{\hat{J}})^2 = \hat{J}_{y}^2$. Regardless of the value of the phase $\eta$ of the driving field, preparation of the spin coherent state $\ket{\theta, \phi} = \ket{0, 0}$ will lead to the most spin squeezing with this OAT term. This is convenient because the state $\ket{\theta, \phi} = \ket{0, 0}$ can typically be prepared by cooling or by optical pumping, without the need for electromagnetic pulses to rotate the spins. Interestingly, the ability to easily change the axis of the spin squeezing term $(\vec{n}_\eta \cdot \vec{\hat{J}})^2$ by changing $\eta$ can be exploited to increase the spin squeezing to the Heisenberg limit using optimal control techniques \cite{She-13}. We note that in our scheme, this does not require control pulses as in \cite{She-13, Liu-11}, but can be achieved by simply shifting the phase $\eta$ of the spin driving field. 

By substituting the expression for $\chi_\text{eff}$ into the approximation condition $\Omega \gg N \chi_\text{eff}$ that leads to Eq. \ref{eq:Heff2}, the condition becomes \begin{equation} \Omega \gg \frac{N\Delta\bar{\lambda}^2}{\Gamma^2} \approx \frac{N\bar{\lambda}^2}{\Delta}, \label{eq:approxcondsground} \end{equation} where on the right hand side we have used the approximation $\Gamma \approx \Delta$, which is valid in the OAT regime. Since (from Eq. \ref{eq:approxconds}) we also have $\Gamma \approx \Delta \gg N \bar{\lambda}$, the condition Eq. \ref{eq:approxcondsground} is easily satisfied for spin drive $\Omega$ comparable to (or greater than) the coupling strength $\bar{\lambda}$. We note, however, that although $\Omega$ should be large we must maintain $ \Gamma\approx\Delta \gg \Omega $ to ensure consistency with our previous approximation conditions in Eq. \ref{eq:approxconds}. 

We now present some numerical results. Neglecting inhomogeneous broadening and inhomogeneous couplings, we plot the minimum spin squeezing $\min_t \xi^2$ as a function of the experimentally adjustable parameters $\Delta$ and $\Omega$ for the initial state $\ket{\theta, \phi} = \ket{0, 0}$ and for a realistic amount of ancillary system relaxation $\gamma$. This is shown for the FQ and NV model in Fig. \ref{fig:OAT_limit_ground}(a). The correspoding plot for the MR model is not shown as it is qualitatively similar. We see in Fig. \ref{fig:OAT_limit_ground}(a) that there are a wide range of values of $\Delta$ and $\Omega$ that give significant spin squeezing. In Fig. \ref{fig:OAT_limit_ground}(b) we plot the time evolution of the spin squeezing parameter $\xi^2$ for several choices of the detuning $\Delta$ and the flux qubit drive $\Omega$, again for the FQ and NV model parameters. Interestingly, even if the condition $\Delta \gg\bar{\lambda}N$ is not well-satisfied, e.g. for $\Delta = 2N\bar{\lambda}$, it is still possible to achieve a level of spin squeezing that is comparable to the optimal squeezing (the horizontal dotted black line in Fig. \ref{fig:OAT_limit_ground}(b)). This is somewhat surprising since we expect such a decrease in $\Delta$ to damage the spin squeezing, as in Fig. \ref{fig:OAT_limit_plus}(c). However, since the spin squeezing is not degraded, it is preferable to choose the smaller value $\Delta =2\bar{\lambda}N$ as the squeezing dynamics are faster in this case.

With this in mind, we estimate the optimum squeezing time for the FQ and NV implementation, with $\Delta = 2\bar{\lambda}N$, $N = 500$ and $\bar{\lambda} = 2\pi\times 12 \text{ kHz}$. We find that $t_\text{opt} = 500 \, \mu\text{s}$. This is a factor of two longer than the corresponding time for the OAT dynamics in section \ref{sec:OAT_plus_ideal} because the OAT coefficient $\chi_\text{eff}/2$ in $\hat{H}'_\text{eff}$ (Eq. \ref{eq:Heff2}) is a factor of two smaller than the OAT coefficient in $\hat{H}_\text{eff}$ (Eq. \ref{eq:Heff}). Similarly, in the microwave resonator model, the optimum squeezing time for $N=1.2\times 10^4$ spins and detuning $\Delta = 2\bar{\lambda}N$ is estimated to be $t_\text{opt} = 320 \text{ ms}$.

\begin{figure}[!ht]
\centering
    \includegraphics[height=45mm]{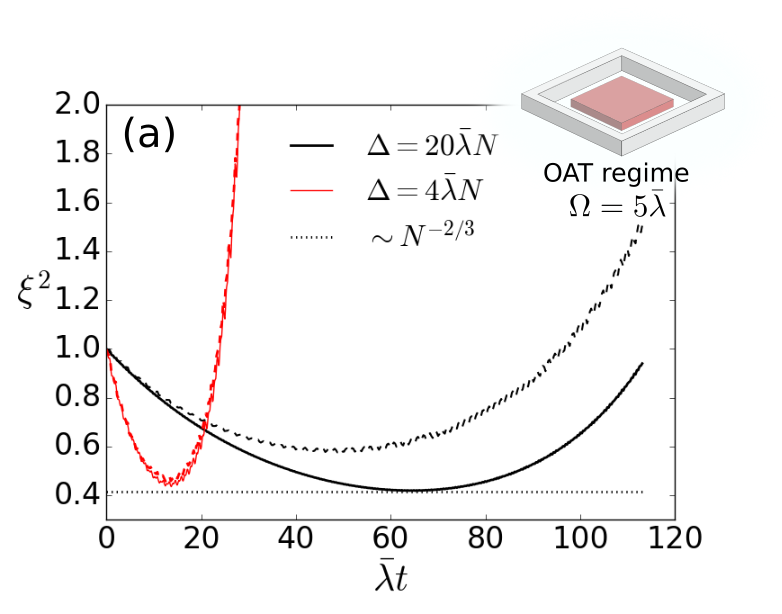}
    \includegraphics[height=45mm]{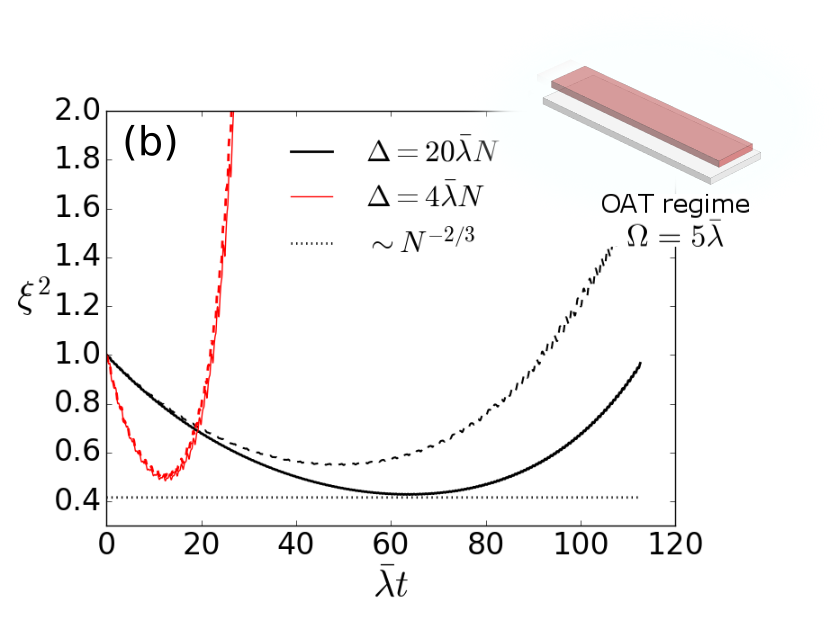}
    \caption{Spin squeezing including inhomogeneous broadening and inhomgogeneous couplings (the dotted lines) for (a) the FQ and NV implemementation ($\delta\lambda = 2\pi\times 1 \text{ kHz}$, $\delta\omega = 2\pi\times 3 \text{ kHz}$), and (b) the MR and donor spins implementation ($\delta\lambda = 2\pi\times 4 \text{ Hz}$, $\delta\omega = 2\pi\times 15 \text{ Hz}$). For comparison, the solid lines show spin squeezing when there is no inhomogeneity ($\delta\lambda = 0,\, \delta\omega = 0$). The effect of inhomogeneous broadening is significantly suppressed due to the spin drive $\Omega\gg\delta\omega$. In Fig. (a), $\bar{\lambda} =2\pi\times 12$ kHz and $\gamma = 0.0265\times\bar{\lambda}N$, and in Fig. (b), $\bar{\lambda} =2\pi\times 56$ Hz and $\gamma = 0.1\times\bar{\lambda}N$. Both (a) and (b) are plotted for $N=6$ and initial state $\ket{\theta,\phi} = \ket{0,0}$ and the dotted lines are averaged over 100 evolutions to remove random fluctuations. Horizontal dotted lines show the optimum spin squeezing by OAT.}    
\label{fig:OAT_limit_ground_inh}
\end{figure} 

\begin{figure*}[!ht]
\centering
    \includegraphics[height=50mm]{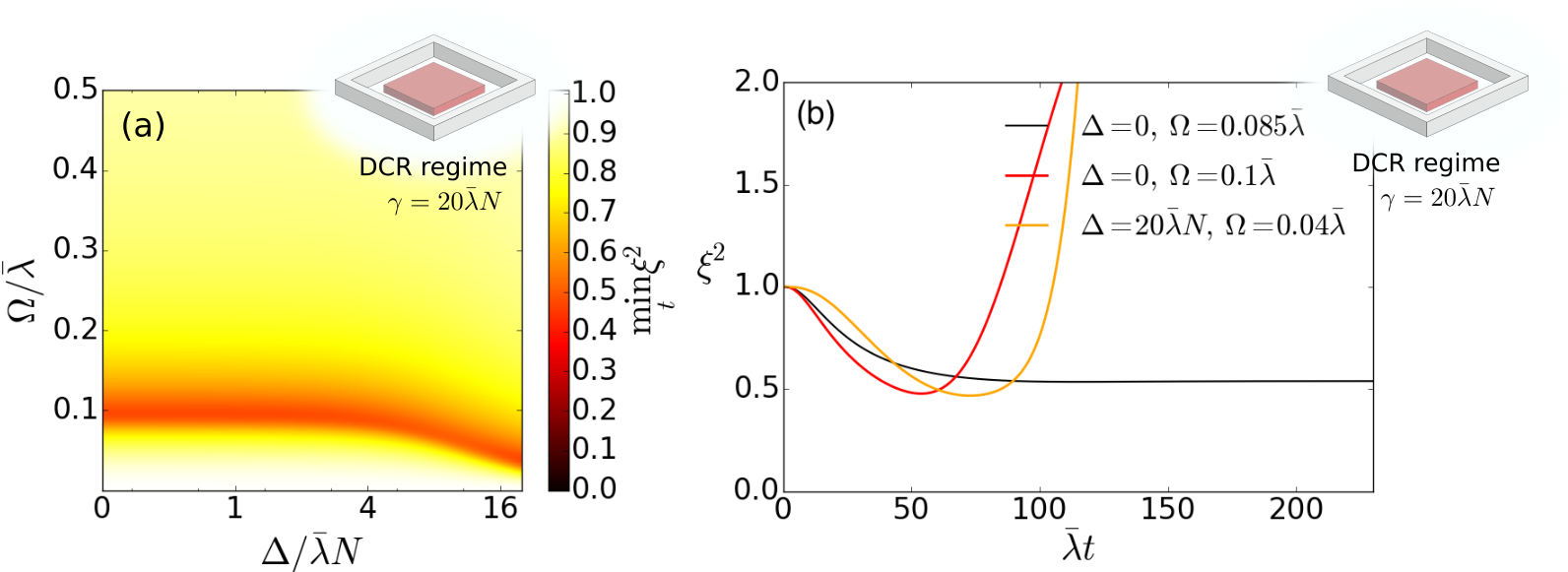}
    \caption{Spin squeezing in the DCR regime ($\Delta \ll \gamma = 20\bar{\lambda}N$). (a) The optimum squeezing that can be achieved for various values of the detuning $\Delta$ and drive $\Omega$. (b) A carefully chosen $\Omega$ leads to steady state spin squeezing (the black line). Both figures were plotted for the flux qubit ancillary system using master Eq. \ref{eq:me1} and Hamiltonian Eq. \ref{eq:Hrwa}, with $N=40$, $\bar{\lambda}=2\pi\times 12$ kHz, $\delta\lambda = 0$, $\delta\omega = 0$, $\gamma = 20\bar{\lambda}N$, and initial state $\ket{\theta,\phi}=\ket{0,0}$.}    
\label{fig:DCR_limit_ground}
\end{figure*} 

Finally, we consider inhomogeneous broadening and inhomogeneous couplings. If, in addition to Eq. \ref{eq:approxcondsground}, we have $\Omega \gg \delta\omega$, the inhomogeneous broadening Hamiltonian in the interaction frame, $\hat{U}^{\prime \dagger}(t) \hat{H}_\text{IB} \hat{U}^{\prime}(t)$, is quickly oscillating and is suppressed in the rotating wave approximation. This is plotted for the FQ and NV device in Fig. \ref{fig:OAT_limit_ground_inh}(a) where the dashed lines show the spin squeezing for $N=6$ spins interacting with a dissipative flux qubit with $\delta\lambda = 2\pi\times 1 \text{ kHz}$ and $\delta\omega = 2\pi\times 3 \text{ kHz}$. For $\Delta = 20\bar{\lambda}N$ (the dashed black line), the spin squeezing is slightly degraded compared to the ideal spin squeezing (the solid black line). This is due to higher order inhomogeneous terms that are not suppressed by the spin drive. However, for smaller detuning $\Delta = 4\bar{\lambda}N$ the spin squeezing is achieved more quickly and inhomogeneous broadening is almost completely suppressed (the dashed red line). Fig. \ref{fig:OAT_limit_ground_inh}(b) shows the corresponding plot for the MR and donor spins implementation.

\section{Spin squeezing by DCR}\label{sec:DCR}
In this section we consider spin squeezing in the DCR regime of our effective master Eq. \ref{eq:effdyn}, that is, when $\Delta \ll \gamma \approx \Gamma$. To easily access this parameter regime we assume that the ancillary system is highly dissipative with $\gamma = 20 \bar{\lambda} N$. We note that this value of ancillary system relaxation is almost three orders of magnitude stronger than the value $\gamma = 0.0265 \times \bar{\lambda} N$ that we used in the previous section for the FQ and NV model, and 200 times larger than $\gamma = 0.1 \bar{\lambda} N$ for the MR and donor spins model.

\begin{figure}[!htbp]
\centering
    \includegraphics[height=50mm]{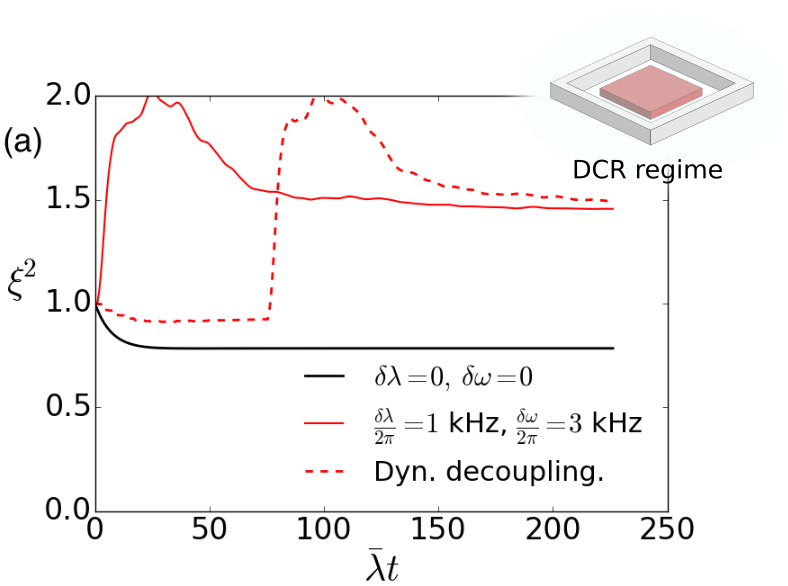}
    \includegraphics[height=50mm]{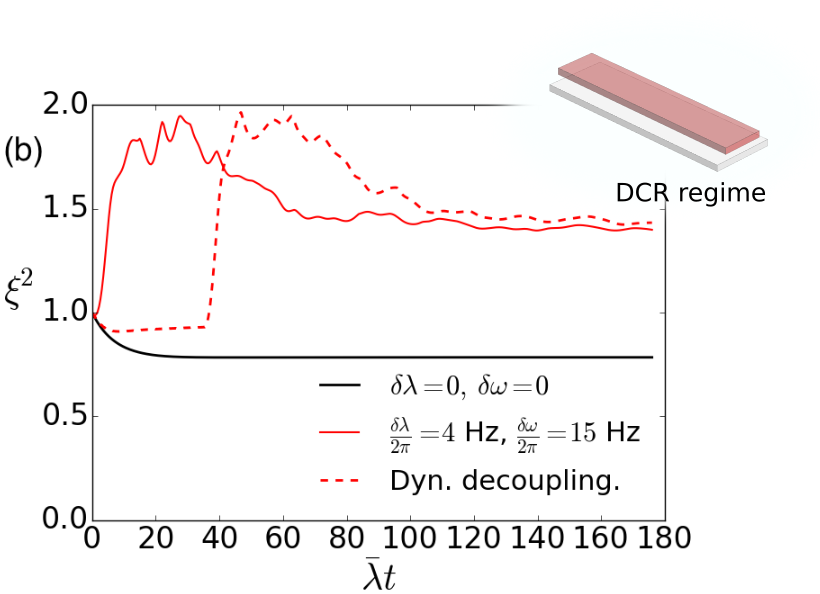}
    \caption{For both (a) the FQ and NV implementation, and (b) the MR and donor spins implementation, the inhomogeneous broadening destroys spin squeezing (the solid red lines) but can be countered with a sequence of operations $\hat{R}S_{\pi}$--$\hat{R}^{\dagger}S_{\pi}$--$\hat{R}S_{\pi}$--$\hat{R}^{\dagger}S_{\pi}$--$...$, where each operation is composed of a rotation $\hat{R} = \hat{R}(-2\theta_\text{ss},\eta)$ (or its inverse $\hat{R}^{\dagger}$) that mimics a reflection of the spin state, and a shift $\eta\to\eta + \pi$ in the phase of the spin drive, represented by $S_{\pi}$. In the dotted red lines we have applied 100 such operations with a free evolution time $\tau$ between them. Each line in the figures above is averaged over 100 evolutions to remove random fluctuations. (For Fig. (a), $N=6$, $\bar{\lambda}=2\pi\times 12$ kHz, $\Delta = 0$, $\Omega = 0.07\bar{\lambda}$, $\gamma = 20\bar{\lambda}N$, $\tau = 0.01 \text{ ms}$, and the initial state is $\ket{\theta,\phi}=\ket{\theta_\text{ss},\phi_\text{ss}}$. For Fig. (b), $N=6$, $\bar{\lambda}=2\pi\times 56$ Hz, $\Delta = 0$, $\Omega = 0.07\bar{\lambda}$, $\gamma = 20\bar{\lambda}N$, $\tau = 1 \text{ ms}$, and the initial state is $\ket{\theta,\phi}=\ket{\theta_\text{ss},\phi_\text{ss}}$.)}    
 \label{fig:N6_DCR}
\end{figure}

\subsection{Ideal case}
Again, we begin by neglecting inhomogeneous broadening and inhomogeneous couplings. In Fig. \ref{fig:DCR_limit_ground}(a), for the FQ and NV model, we plot $\min_t\xi^2$ as a function of the detuning $\Delta$ and the spin drive $\Omega$ for the easily prepared initial state $\ket{\theta,\phi}=\ket{0,0}$. We see that for a range of values of $\Delta$ and $\Omega$ there is significant spin squeezing during the evolution. In Fig. \ref{fig:DCR_limit_ground}(b) we plot the time evolution of the spin squeezing for various choices of $\Delta$ and $\Omega$. As expected \cite{Gon-13,Wol-14}, we see steady state spin squeezing for a carefully chosen value of the spin drive $\Omega$ (the black line, Fig. \ref{fig:DCR_limit_ground}(b)). We have verified numerically that for these parameters the steady state of the master Eq. \ref{eq:me1} with the Hamiltonian Eq. \ref{eq:Hrwa} is indeed squeezed. 

\subsection{Realistic case, dynamical decoupling}


We now consider the effect of inhomogeneous broadening and inhomogeneous couplings on DCR spin squeezing. In this case our numerics are limited to a small number of spins $N=6$. For simplicity we also assume that $\Delta = 0$, i.e., the ancillary system and the spin system are resonant, so that the effective Hamiltonian Eq. \ref{eq:Heff} only includes the spin drive term $\hat{H}_\text{eff} =  \hbar\Omega\, \vec{n}_\eta \cdot \vec{\hat{J}}$, and the effective master equation is of the form Eq. \ref{eq:meCR}. For the FQ and NV model, the red line in Fig. \ref{fig:N6_DCR}(a) shows that for $\delta\omega = 2\pi\times 3 \text{ kHz}$ and $\delta\lambda = 2\pi\times 1 \text{ kHz}$ the spin squeezing is quickly degraded. The red line in Fig. \ref{fig:N6_DCR}(b) shows that the inhomogeneities have a similar effect on spin squeezing for the parameters in the MR model. Further numerics have shown that, as with OAT, this is primarily due to the inhomogeneous broadening $\hat{H}_\text{IB}$. Unfortunately, the dynamical decoupling approach that was taken to protect spin squeezing against inhomogenous broadening in the previous section for OAT will not work for DCR. This is because for spin squeezing by OAT, the $\pi$-pulse operator $\hat{R}(\pi,\phi) = \exp[-i\pi(\hat{J}_x\sin\phi -\hat{J}_y\cos\phi)]$ has the convenient property that it commutes with the OAT operator $\hat{J}_z^2$ so that $\hat{R}^{\dagger}(\pi,\phi)\hat{J}_z^2 \hat{R}(\pi,\phi) = \hat{J}_z^2$ and the spin squeezing is not disrupted by the $\pi$-pulse. For DCR, however, the $\pi$-pulse operator $\hat{R}(\pi,\phi)$ will disrupt the DCR squeezing mechanism. To see this we start from the effective master Eq. \ref{eq:effdyn} in the DCR regime $\Delta \ll \gamma$ (assuming $\Delta = 0$ and $\delta\lambda = 0$): \begin{equation} \dot{\rho}_\text{s} = -\frac{i}{\hbar} [\hat{H}_\text{IB} + \hbar\Omega\,\vec{n}_\eta\cdot\vec{\hat{J}} , \rho_\text{s}] + \gamma_\text{eff}\mathcal{D}[\hat{J}_{-}](\rho_\text{s}) . \label{eq:DCR_t<tau} \end{equation} If at time $\tau$ we apply the $\pi$-pulse operator $\hat{R}(\pi,\phi)$, the state is transformed to $\rho'_\text{s}(\tau) = \hat{R}(\pi,\phi)\rho_\text{s}(\tau)\hat{R}^\dagger(\pi,\phi)$ and the master equation for the following period of time $t>\tau$ is: \begin{equation} \dot{\rho}'_\text{s} = -\frac{i}{\hbar} [\hat{H}_\text{IB} + \hbar\Omega\,\vec{n}_\eta\cdot\vec{\hat{J}} , \rho'_\text{s}] + \gamma_\text{eff}\mathcal{D}[\hat{J}_{-}](\rho'_\text{s}) . \label{eq:DCR_t>tau_1} \end{equation} Operating on Eq. \ref{eq:DCR_t>tau_1} on the left by $\hat{R}^\dagger(\pi,\phi)$ and on the right by $\hat{R}(\pi,\phi)$ gives (for $t>\tau$) the evolution equation: \begin{equation} \dot{\rho}_\text{s} = -\frac{i}{\hbar} [ - \hat{H}_\text{IB} + \hbar\Omega\hat{R}^\dagger(\vec{n}_\eta\cdot\vec{\hat{J}})\hat{R} , \rho_\text{s}] + \gamma_\text{eff}\mathcal{D}[\hat{J}_{+}](\rho_\text{s}) , \label{eq:DCR_t>tau_2} \end{equation} where we have used $\hat{R}^\dagger(\pi,\phi)\hat{H}_\text{IB} \hat{R}(\pi,\phi) = - \hat{H}_\text{IB}$ and $\hat{R}^\dagger(\pi,\phi) \hat{J}_{-} \hat{R}(\pi,\phi) = - e^{-2i\phi} \hat{J}_{+}$. Comparing Eq. \ref{eq:DCR_t<tau} and Eq. \ref{eq:DCR_t>tau_2} shows that the effect of the $\pi$-pulse is to reverse the sign of the inhomogeneous broadening Hamiltonian $\hat{H}_\text{IB}$ in the following period of evolution. However, unlike for OAT, the DCR spin squeezing mechanism is also disrupted by the $\pi$-pulse, since the collective relaxation operator is transformed from $\hat{J}_{-}$ to $\hat{J}_{+}$. For example, if the system was in the steady state of the DCR master equation before the $\pi$-pulse, then after the $\pi$-pulse it will be far from the steady state. 

However, the desired effect can be achieved by a reflection of each spin at time $\tau$ through a plane in the Bloch sphere that contains the $\vec{z}$-axis and the vector $\vec{n}_\eta$. Such a reflection is implemented by the complex conjugation operator $\hat{V}$, with the assumption that for each spin the matrix elements of $\hat{\sigma}_z^{(i)}$, $\vec{n}_\eta\cdot\vec{\hat{\sigma}}^{(i)}$ and $\vec{n}_{\eta+\pi/2}\cdot\vec{\hat{\sigma}}^{(i)}$ are: \begin{eqnarray} \hat{\sigma}_z^{(i)} = \left( \begin{array}{cc} 1 & 0 \\ 0 & -1 \\ \end{array} \right) , \\ \vec{n}_\eta\cdot\vec{\hat{\sigma}}^{(i)} = \left( \begin{array}{cc} 0 & 1 \\ 1 & 0 \\ \end{array} \right), \\ \vec{n}_{\eta+\pi/2}\cdot\vec{\hat{\sigma}}^{(i)} = \left( \begin{array}{cc} 0 & -i \\ i & 0 \\ \end{array} \right) . \end{eqnarray} This gives: \begin{eqnarray} \hat{V}\hat{\sigma}_z^{(i)} \hat{V}^{-1} &=& \hat{\sigma}_z^{(i)}, \nonumber\\ \hat{V}(\vec{n}_\eta\cdot\vec{\hat{\sigma}}^{(i)})\hat{V}^{-1} &=& \vec{n}_\eta\cdot\vec{\hat{\sigma}}^{(i)}, \nonumber\\  \hat{V}(\vec{n}_{\eta+\pi/2}\cdot\vec{\hat{\sigma}}^{(i)})\hat{V}^{-1} &=& - \vec{n}_{\eta+\pi/2}\cdot\vec{\hat{\sigma}}^{(i)} . \end{eqnarray} Applying the complex conjugation operator to the spin state at time $\tau$ transforms the state to $\rho'_\text{s}(\tau) = \hat{V}\rho_\text{s}(\tau)\hat{V}^{-1}$. The master equation for the following period of evolution is: \begin{equation} \dot{\rho}'_\text{s} = -\frac{i}{\hbar} [\hat{H}_\text{IB} + \hbar\Omega\,\vec{n}_\eta\cdot\vec{\hat{J}} , \rho'_\text{s}] + \gamma_\text{eff}\mathcal{D}[\hat{J}_{-}](\rho'_\text{s}) . \label{eq:DCR_t>tau_3} \end{equation} Operating on Eq. \ref{eq:DCR_t>tau_3} on the left by $\hat{V}^{-1}$ and on the right by $\hat{V}$ gives, for $t>\tau$, the evolution equation: \begin{equation} \dot{\rho}_\text{s} = -\frac{i}{\hbar} [ - \hat{H}_\text{IB} - \hbar\Omega\,\vec{n}_\eta\cdot\vec{\hat{J}} , \rho_\text{s}] + \gamma_\text{eff}\mathcal{D}[\hat{J}_{-}](\rho_\text{s}) , \label{eq:DCR_t>tau_4} \end{equation} where we have used $\hat{V}^{-1}i\hat{V} = -i$, $\hat{V}^{-1}\hat{H}_\text{IB}\hat{V} = \hat{H}_\text{IB}$, $\hat{V}^{-1} (\vec{n}_\eta\cdot\vec{\hat{J}}) \hat{V} = \vec{n}_\eta\cdot\vec{\hat{J}}$ and $\hat{V}^{-1}\hat{J}_{-}\hat{V} = \hat{J}_{-}$. Comparing Eq. \ref{eq:DCR_t<tau} and Eq. \ref{eq:DCR_t>tau_4} we see that the sign of the inhomogeneous broadening Hamiltonian $\hat{H}_\text{IB}$ is reversed and the Lindblad term is unchanged, as desired. However, the spin drive term is also transformed from $\Omega \,\vec{n}_\eta\cdot\vec{\hat{J}}$ to $-\Omega \,\vec{n}_\eta\cdot\vec{\hat{J}}$. This can easily be corrected by shifting the phase of the spin drive $\eta \to \eta + \pi$ so that $-\Omega \,\vec{n}_\eta\cdot\vec{\hat{J}} \to \Omega \,\vec{n}_\eta\cdot\vec{\hat{J}}$, which finally gives: \begin{equation} \dot{\rho}_\text{s} = -\frac{i}{\hbar} [ - \hat{H}_\text{IB} + \hbar\Omega\,\vec{n}_\eta\cdot\vec{\hat{J}} , \rho_\text{s}] + \gamma_\text{eff}\mathcal{D}[\hat{J}_{-}](\rho_\text{s}) . \label{eq:DCR_t>tau_5} \end{equation} Eq. \ref{eq:DCR_t>tau_5} is identical to Eq. \ref{eq:DCR_t<tau}, apart from a reversal of the inhomogeneous broadening. We note that this operation must be repeated many times, with a short free evolution time $\tau$ between each operation, since the dynamics before the reflection and phase-shift does not commute with the dynamics afterwards. 

Finally, we note that the reflection is an unphysical transformation, since the complex conjugation operator is anti-unitary. However, for some states, we can apply rotations that have the same effect as a reflection. For example, reflecting each spin of the spin coherent state $\ket{\theta,\pi/2}$ in the $xz$-plane of its Bloch sphere gives the state $\ket{\theta,-\pi/2}$. Clearly, this transformation can also be implemented by a rotation $\hat{R}(-2\theta,\pi/2) = \exp[i2\theta\hat{J}_x]$ of each spin around the $x$-axis. The angle of rotation $2\theta$ depends on the spin coherent state parameter $\theta$. For simplicity, we assume that the spin system is prepared in a spin coherent state that is ``close'' to the steady state. We take this to be the state $\ket{\theta_\text{ss},\phi_\text{ss}}$ where the angle $\theta_\text{ss} = \cos^{-1}(-2\langle\hat{J}_z\rangle_\text{ss}/N)$ is determined by the expectation value $\langle \hat{J}_z \rangle_\text{ss}$ in the steady state and $\phi_\text{ss} = \tan^{-1}(\langle \hat{J}_y \rangle_\text{ss}/ \langle \hat{J}_x \rangle_\text{ss}) = \eta + \frac{\pi}{2}$. This simplifies the procedure because the rotation that mimics the reflection of the state does not have to be changed as the system evolves. The rotation is $\hat{R}(-2\theta_\text{ss},\eta)$, which, for example, transforms the state $\ket{\theta_\text{ss}, \eta + \frac{\pi}{2} }$ to $\ket{\theta_\text{ss} ,  \eta - \frac{\pi}{2} }$, the reflection of the state through the plane that contains the $\vec{z}$-axis and the vector $\vec{n}_\eta$. The result of repeating this operation many times is plotted in the dashed red lines, Fig. \ref{fig:N6_DCR}. We see that the inhomogeneous broadening can be significantly suppressed by this procedure and that some spin squeezing can be recovered. We note that it may be possible to get a further improvement by alternating reflections of the state in the plane containing the $\vec{z}$-axis and $\vec{n}_\eta$ with those in the orthogonal plane containing the $\vec{z}$-axis and $\vec{n}_\eta\times \vec{z}$. This would be analagous to the alternating $\pi$-pulses around two orthogonal axes in the concatenated-XY8 pulse sequence in section \ref{sec:OAT_dd}.

\section{Conclusion}
We have shown that a single model -- the interaction of a spin system with a dissipative ancillary system -- can lead to spin squeezing by two distinct mechanisms: one-axis twisting and driven collective relaxation. In either case, spin squeezing is generated even though the ancillary system coherence time is much smaller than the duration of the squeezing process. This is possible because we have adiabatically eliminated the ancillary system, which stays close to its ground state throughout the squeezing process. We focus on two possible implementations, with either a superconducting flux qubit or a superconducting microwave resonator playing the role of the ancillary system. With dynamical decoupling we have shown numerically that both spin squeezing mechanims are robust to inhomogeneities in the model. In practice, the dynamical decoupling pulses may introduce errors that reduce the spin squeezing. However, we have also shown that by driving the spin system it is possible to generate robust OAT spin squeezing without the need for electromagnetic pulses. This flexibility -- squeezing can be generated in disparate parameter regimes and for a variety of practical requirements -- is, we believe, a strength of this model. We conclude that spin squeezing of hundreds of solid state electron spins should be experimentally feasible in this model with current or near future technology.

\begin{figure}[!htbp]
\centering
    \includegraphics[height=50mm]{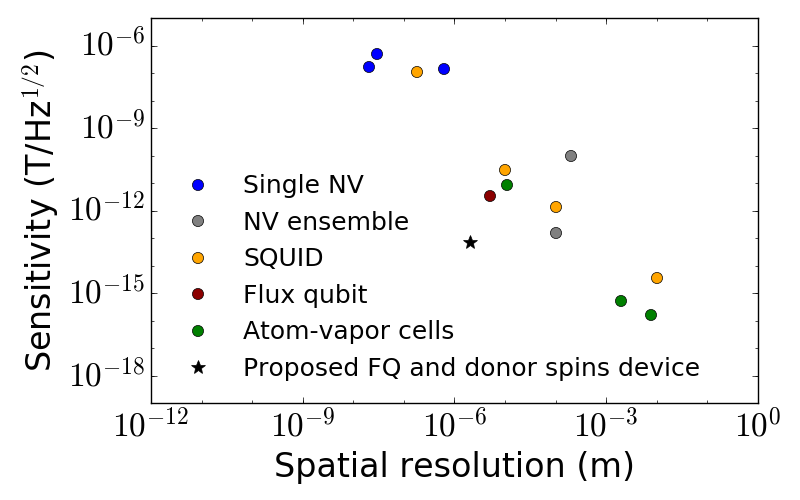}
    \caption{Comparison of the proposed spin squeezed magnetometer (the black star; based on a flux qubit and donor spins implementation) with reported values for existing devices \cite{Maz-08, Pha-11, Mal-12, Aco-09, LeS-12, Kir-95, Bau-03, Fal-06, Fin-10, Bal-12, Kom-03, Ven-07, Dan-10}.}    
 \label{fig:s_o_t_a}
\end{figure}

Finally, we estimate the sensitivity of a magnetic field measurement that can be achieved using a spin system prepared in a squeezed state. We assume that the squeezed state is prepared by OAT, since this leads to more spin squeezing than DCR. We also assume that the spin system is undergoing non-Markovian dephasing during the field sensing period. In this case, using the recent results of \cite{Tan-15}, we estimate that for our FQ and NV implementation a magnetic field $B$ can be measured with sensitivity $\delta B \sqrt{T} = 1.4 \text{ pT}/\sqrt{\text{Hz}}$ where $T$ is the total sensing time (see Appendix \ref{app:Bsensing} for details). This is a factor of $\sim 2.7$ times improvement over the best sensitivity that can be achieved using a separable state of the spins. Such a magnetic field sensor would also have a very good spatial resolution $\sim 2 \, \mu\text{m}$, as determined by the size of the diamond sample. For the MR and donor spin implementation, we estimate $\delta B \sqrt{T} = 10 \text{ fT}/\sqrt{\text{Hz}}$, a factor of $\sim 4.1$ improvement over the best sensitivity that can be achieved using a separable state of the donor spins (see Appendix \ref{app:Bsensing} for details), and a spatial resolution $\sim 1 \text{ mm}$. There is better sensitivity for the MR implementation than for the FQ implementation since it employs a higher number of spins, and because these spins are donor electrons in silicon, which have much longer coherence times than NV centres \cite{Tyr-12}. However, the spatial resolution is worse since the sample coupled to the MR is larger than the sample coupled to the FQ (see Fig. \ref{fig:model}). By using a flux qubit ancillary system with $N=500$ donor spins in silicon we could combine the best features of both implementations, giving a sensitivity $\delta B \sqrt{T} = 75 \text{ fT}/\sqrt{\text{Hz}}$ and a spatial resolution $\sim 2\,\mu\text{m}$. In Fig. \ref{fig:s_o_t_a} we show how such a magnetometer would compare with the reported sensitivies of some existing state-of-the-art magnetometers. Relative to these existing devices, our proposed magnetometer would occupy an unexplored region of high sensitivity and high spatial resolution. The sensitivity, and the advantage of using a squeezed state instead of a separable state, can be improved by generating a squeezed state with a larger number of spins. To do this, the challenge is to couple a larger number of spins to the ancillary system, while maintaining long coherence times and strong, relatively homogeneous coupling. We note that this should be possible as experimental techniques advance. For example, if a superconducting circuit can be arranged in a Helmholz coil configuration, that is, with two superconducting solenoids separated by the solenoid radius, then the coupling to a spin ensemble placed between the solenoids will be stronger, and more homogeneous over a larger spatial region than for the setups in Fig. \ref{fig:model}. This would enable the generation of much larger spin squeezed states.

\begin{acknowledgments}
We would like to thank Dr. Kousuke Kakuyanagi, Dr. Hikaru Toida, and Dr. Yuimaru Kubo for valuable discussions about the realistic parameters available in current experiments. We also acknowledge support from the MEXT Grant-in-Aid for Scientific Research on Innovative Areas ``Science of hybrid quantum systems'' Grant number 15H05870, the MEXT Grant-in-Aid for Scientific Research(S) Grant number 25220601, and the JSPS KAKENHI Grant number 15K17732. S.D. acknowledges support from the Japan Society for the Promotion of Science (JSPS). Finally, the numerical simulations used the QuTiP package \cite{Joh-12,Joh-13}.
\end{acknowledgments}

\bibliography{refs}

\appendix

\section{Derivation of effective master equation}\label{app:eff_me_deriv}
To derive the effective master Eq. \ref{eq:effdyn} we follow the procedure developed by Reiter and Sorensen \cite{Rei-12} for adiabatic elimination in an open quantum system. The starting point is the master Eq. \ref{eq:me1}, $\dot{\rho} = - \frac{i}{\hbar} [\hat{H}, \rho] + \gamma\mathcal{D}[\hat{A}](\rho)$, where the Lindblad operator $\hat{A} = \sum_{n=0}^{d-2}\sqrt{n+1}\ket{n}\bra{n+1}$ represents dissipation of energy into the ground state of the ancillary system. The projector onto the ancillary system ground state is denoted $\hat{P}_g = \ket{0}\bra{0}$, while $\hat{P}_e = \mathbb{I} - \hat{P}_g = \sum_{n=1}^{d-1}\ket{n}\bra{n}$ projects onto the excited subspace. This allows the Hamiltonian to be written as $\hat{H} = (\hat{P}_g + \hat{P}_e) \hat{H} (\hat{P}_g + \hat{P}_e) = \hat{V}_g + \hat{V}_e + \hat{V}_{+} + \hat{V}_{-}$, where \begin{eqnarray} \hat{V}_g &=& \hat{P}_g \hat{H} \hat{P}_g = (\hbar\Omega\vec{n}_{\eta}\cdot \vec{\hat{J}} + \hat{H}_\text{IB} ) \otimes \hat{P}_g , \nonumber \\ \hat{V}_e &=& \hat{P}_e \hat{H} \hat{P}_e = (\hbar\Omega\vec{n}_{\eta}\cdot \vec{\hat{J}} + \hat{H}_\text{IB} ) \otimes \hat{P}_e + \hbar\Delta\sum_{n=1}^{d-1}n\ket{n}\bra{n} , \nonumber \end{eqnarray} describe the dynamics within the ground and excited subspaces respectively, while \begin{eqnarray} \hat{V}_+ &=& \hat{P}_e \hat{H} \hat{P}_g = \hbar\bar{\lambda} \hat{J}_- \otimes \ket{1}\bra{0} , \nonumber \\ \hat{V}_- &=& \hat{P}_g \hat{H} \hat{P}_e = \hbar\bar{\lambda} \hat{J}_+ \otimes \ket{0}\bra{1} , \nonumber \end{eqnarray} give the dynamics that connects the two subspaces. 

The Reiter-Sorensen procedure gives a prescription for the derivation of an effective master equation under the assumptions that the dynamics due to $\hat{V}_g$ and $\hat{V}_{\pm}$ are much slower than either the dissipative dynamics or the dynamics due to $\hat{V}_e$. In our model this requirement is satisfied if $\Gamma \gg \bar{\lambda}N$, $\Gamma \gg \Omega$ and $\Gamma \gg \delta\omega$, where $\Gamma = \sqrt{\Delta^2 + \gamma^2/4}$. According to the Reiter-Sorensen procedure, the resulting effective master equation is $\dot{\rho} = -\frac{i}{\hbar}[\hat{V}_\text{eff},\rho] + \mathcal{D}[\hat{L}_\text{eff}](\rho)$ with effective Hamiltonian and Lindblad operators \begin{eqnarray} \hat{V}_\text{eff} &=& -\frac{1}{2} \hat{V}_{-} [ \hat{V}_\text{NH}^{-1} + (\hat{V}_\text{NH}^{-1})^{\dagger}] \hat{V}_{+} + \hat{V}_g , \label{eq:Veff} \\ \hat{L}_\text{eff} &=& \sqrt{\gamma}\hat{A} \hat{V}_\text{NH}^{-1}\hat{V}_{+} , \label{eq:Leff} \end{eqnarray} where \begin{eqnarray} \hat{V}_\text{NH} &=& \hat{V}_e - \frac{i\hbar\gamma}{2}\hat{A}^{\dagger}\hat{A} \label{eq:VNH} \\ &=& (\hbar \Omega \vec{n}_{\eta}\cdot \vec{\hat{J}} + \hat{H}_\text{IB} ) \otimes \hat{P}_e + \hbar\left(\Delta - \frac{i\gamma}{2} \right) \sum_{n=1}^{d-1}n\ket{n}\bra{n} . \nonumber \end{eqnarray} We refer the reader to Ref. \cite{Rei-12} for a derivation of the effective operators Eqs. \ref{eq:Veff}, \ref{eq:Leff} and \ref{eq:VNH}. (The essence of the approximation is that the dynamics due to $\hat{V}_g$ and $\hat{V}_{\pm}$ are perturbatively small compared to the dynamics due to the (non-Hermitian) Hamiltonian $\hat{V}_\text{NH}$, allowing adiabatic elimination of the excited subspace.)


Since we have already assumed that $\Gamma \gg \max\{ \Omega, \delta\omega \}$, we can approximate \begin{eqnarray} \hat{V}_\text{NH}^{-1} &\approx& \frac{1}{\hbar (\Delta -i\gamma /2)} \sum_{n=1}^{d-1}\frac{1}{n}\ket{n}\bra{n} \\  &=& \frac{\Delta +i\gamma /2}{\hbar \Gamma^2} \sum_{n=1}^{d-1}\frac{1}{n}\ket{n}\bra{n} . \label{eq:VNHinv} \end{eqnarray} Substituting Eq. \ref{eq:VNHinv} into Eqs. \ref{eq:Veff} and \ref{eq:Leff} gives: \begin{eqnarray} \hat{V}_\text{eff} &=& \left( \hbar \Omega \vec{n}_{\eta}\cdot \vec{\hat{J}} + \hat{H}_\text{IB} - \frac{\hbar\Delta\lambda^2}{\Gamma^2} \hat{J}_{+}\hat{J}_{-} \right) \otimes \hat{P}_g , \label{eq:VeffFinal} \\  \hat{L}_\text{eff} &=& \frac{\sqrt{\gamma}\lambda (\Delta + i\gamma/2)}{\Gamma^2}\hat{J}_- \otimes \hat{P}_g . \label{eq:LeffFinal} \end{eqnarray} We may ignore the projector $\hat{P}_g$ in Eqs. \ref{eq:VeffFinal} and \ref{eq:LeffFinal} since the ancillary system remains in its ground state throughout the evolution. Using the identity $\hat{J}_{+}\hat{J}_{-} = \vec{\hat{J}} \cdot \vec{\hat{J}} - \hat{J}_z^2 + \hat{J}_z$ we see that the effective Hamiltonian is $\hat{V}_\text{eff} = \hat{H}_\text{IB} + \hat{H}_\text{eff}$ where $\hat{H}_\text{eff}$ is given by Eq. \ref{eq:Heff}. It is also easy to verify that $\mathcal{D}[\hat{L}_\text{eff}](\rho) = \frac{\gamma\lambda^2}{\Gamma^2}\mathcal{D}[\hat{J}_{-}](\rho)$, resulting in the effective master Eq. \ref{eq:effdyn}.

\section{Magnetic field sensing with the spin squeezed state}\label{app:Bsensing}
In Ref. \cite{Tan-15}, it was shown that the sensitivity of magnetic field estimation -- taking non-Markovian dephasing of the spins into account -- is \begin{equation} \delta B \sqrt{T} \approx \frac{\hbar}{g_e \mu_B} \sqrt{ \left( \frac{ t + t_\text{opt} }{t^2} \right) \left( \frac{2\xi^2}{N} + \frac{N(e^{2\gamma_s^2 t^2} - 1)}{2 |\langle \vec{\hat{J}} \rangle|^2} \right) } , \label{eq:sens} \end{equation} where $t$ is the sensing time for each measurement, $T = t\nu$ is the total measurement time ($\nu$ is the number of repetitions of the measurement), $\gamma_\text{s}^{-1}$ is the spin coherence time, $t_\text{opt}$ is the time taken to prepare the spin squeezed state, $\xi^2$ is the Wineland squeezing parameter, $g_e$ is the electron $g$-factor, $\mu_B$ is the Bohr magneton, and the magnetic field estimatate is obtained by making a measurement of a collective spin observable in the direction of least variance \cite{Tan-15}. (Eq. \ref{eq:sens} corresponds to Eq. 43 in the supplementary material of \cite{Tan-15}.)

For the FQ and NV implementation, we assume an NV coherence time $\gamma_s^{-1} = 30 \text{ ms}$ \cite{Far-15} and we use $N=500$, as estimated in section \ref{sec:FQNV}. For magnetic field sensing with a spin coherent state we assume \begin{equation} t_\text{opt} = 0,\quad |\langle \vec{\hat{J}} \rangle| = \frac{N}{2},\quad \xi^2 = 1 . \label{eq:scssens} \end{equation} Substituting into these values into Eq. \ref{eq:sens} means that it is a function of $t$ alone, that can be minimised numerically with respect to $t$ to find the best achievable sensitivity for a separable state, $\delta B\sqrt{T} = 3.8 \text{ pT}/\sqrt{\text{Hz}}$. For an OAT spin squeezed state of the NV centres we use $t_\text{opt} = 250\,\mu\text{s}$, as estimated in section \ref{sec:OAT_plus_ideal}, $|\langle \vec{\hat{J}} \rangle| = (N/2)\cos^{N-1}(\Theta/2)$ \cite{Ma-11} and $\xi^2 = \frac{N^2}{4|\langle \vec{\hat{J}} \rangle|^2} [1 - (N-1)C]$ \cite{Ma-11} where $\Theta = 2\chi_\text{eff}t_\text{opt}$ and: \begin{eqnarray} && C = - \frac{1}{4}(1 - \cos^{N-2}\Theta) + \nonumber \\ && \frac{1}{4}\left[ (1 - \cos^{N-2}\Theta)^2 + 16\sin^2\frac{\Theta}{2}\cos^{2N-4}\frac{\Theta}{2} \right]^{1/2} . \end{eqnarray} After minimising Eq. \ref{eq:sens} with respect to $t$, this gives a sensitivity $\delta B\sqrt{T} = 1.4 \text{ pT}/\sqrt{\text{Hz}}$, a factor of $\sim 2.7$ improvement over a spin coherent state. 

For the MR and donor spins implementation, we assume a spin coherence time $\gamma_\text{s} = 10 \text{ s}$ \cite{Tyr-12} and we use $N = 1.2\times 10^4$, as estimated in section \ref{sec:MRdonor}. For a spin coherent state, using the values in Eq. \ref{eq:scssens} and minimising Eq. \ref{eq:sens} with respect to $t$ gives a sensitivity $\delta B\sqrt{T} = 42 \text{ fT}/\sqrt{\text{Hz}}$. For a squeezed state we obtain $\delta B\sqrt{T} = 10 \text{ fT}/\sqrt{\text{Hz}}$, which is an improvement over the spin coherent state by a factor of $\sim 4.1$. 

Finally, if the NV centres in the FQ implementation are replaced with donor spins in silicon, we can combine the good spatial resolution of the FQ implementation with the good senstivity resulting from the long coherence times of the donor spins. In this case, substituting $N = 500$ and $\gamma_\text{s} = 10 \text{ s}$ gives a sensitivity of $\delta B\sqrt{T} = 75 \text{ fT}/\sqrt{\text{Hz}}$ using an OAT squeezed state, and a spatial resolution $\sim 2 \,\mu\text{m}$.

\end{document}